\documentclass[9pt,twoside]{rmaa-rho-class/rmaa-rho_light}
\RMxAAtemplatetype{\RMxAA} 
\usepackage{fancyhdr}

\vol{}
\pages{}
\thisyear{2026}
\doi{TDB}

\title{Discovery of Ultra Fast Outflows in the Narrow Line Seyfert 1 Galaxy Mrk 110 with \textit{XMM-Newton} and \textit{NuSTAR}}

\author[1]{O. Segura Montero}
\author[1]{A. L. Longinotti}
\author[1]{Y. Krongold}
\author[2]{A. Marinucci}
\author[3]{M. Guainazzi}
\author[4,5]{S. Mathur}
\author[1]{M. Gudi\~no}
\author[1]{C. I. Victoria-Ceballos}
\author[1]{A. Lara-DI}
\author[6]{M. Santos-Lle\'o}

\affil[1]{Universidad Nacional Aut\'onoma de M\'exico. Instituto de Astronom\'ia. A.P. 70-264, 04510. Ciudad de M\'exico, M\'exico.}
\affil[2]{ASI - Agenzia Spaziale Italiana, Via del Politecnico snc, I-00133 Roma, Italy}
\affil[3]{ESA European Space Research and Technology Centre (ESTEC), Keplerlaan 1, NL-2201 AZ Noordwijk, the Netherlands}
\affil[4]{Astronomy Department, The Ohio State University, Columbus, OH 43210, USA}
\affil[5]{Center for Cosmology and Astro-Particle Physics, The Ohio State University, Columbus, OH 43210, USA}
\affil[6]{Formerly: European Space Agency (ESA), European Space Astronomy Centre (ESAC), Camino Bajo del Castillo s/n, E-28692 Villanueva de la Cañada, Madrid, Spain}

\leadauthor{Segura Montero et al.}
\smalltitle{Ultra Fast Outflows on Mrk 110}

\corres{O. Segura Montero}
\email{osegura@astro.unam.mx}

\received{}
\accepted{}

\license{Texto de la licencia aquí}

\setbool{rho-abstract}{true} 
\setbool{rho-resumen}{true} 

\begin{abstract}
    We present an X-ray analysis of {\it XMM-Newton} and {\it NuSTAR} spectra of the Narrow-Line Seyfert 1 (NLS1) galaxy Mrk 110. Our analysis reveals the intriguing presence of a multi-component ultra fast outflow (UFO) both in the grating spectrum and the NuSTAR data. The moderate UFO significance (3-4$\sigma$), in the grating and CCD data indicates a shallow absorber, consistently with the absence of previous such claims in this source. The two components (UFO1 and UFO2) in the grating spectrum show a fairly wide range of velocities (0.05-0.06c), as their respective ionization state (Log U = -1.43 and -1.82), however they share the same column density (Log N$_H$ $\sim$ 19.7). In the {\it NuSTAR} spectrum, we found a much faster  wind component with a velocity of $\sim$ 0.21c (UFO3). We show that, in an energy-driven wind scenario, UFO3 might carry enough energy to produce significant feedback on its host galaxy. Considering its known radio properties, Mrk 110 is therefore another Narrow-Line Seyfert 1 with stratified ultra-fast outflows co-existent with a pc-scale jet/outflow radio structure.
\end{abstract}


\begin{resumen}
    Presentamos un an\'alisis de rayos X de los espectros {\it XMM-Newton} y {\it NuSTAR} de la galaxia Narrow-Line Seyfert 1 Mrk 110. Nuestro an\'alisis revela la presencia de un flujo de salida r\'apido de m\'ultiples componentes tanto en el espectro de alta resoluci\'on como en los datos de NuSTAR. La significancia de los componentes en los datos de alta resoluci\'on y CCD es 3 y 4$\sigma$, que consisten con la presencia de un absorbente superficial, previamente no visto en esta fuente. Los dos componentes (UFO1 y UFO2) en el espectro RGS muestran un rango de velocidades de 0,05c y 0,06c, aunque su densidad de columna (Log N$_H$ $\sim$ 19.7) y estado de ionizaci\'on (Log U = -1.43 y -1.82) son bastante similares. En el espectro {\it NuSTAR}, encontramos un componente a\'un m\'as r\'apida con una velocidad de $\sim$ 0.21c. Mostramos que, en un escenario de viento impulsado por energ\'ia, UFO3 podr\'ia transportar suficiente energ\'ia para producir una retroalimentaci\'on significativa en su galaxia anfitriona. Considerando sus propiedades de radio conocidas, Mrk 110 es por lo tanto otro Seyfert 1 de l\'inea estrecha con flujos de salida ultra-r\'apidos estratificados coexistentes con una estructura de radio de chorro/flujo de salida a escala de pc.
\end{resumen}

\begin{document}
\nolinenumbers
\maketitle
\pagestyle{fancy}\thispagestyle{firststyle}


\section{INTRODUCTION}

\label{sec:intro}
Active Galactic Nuclei (AGN) are very luminous compact regions observed  at the center of galaxies, whose extremely high luminosity output is  not directly attributed to stars. They are powered by supermassive black holes (SMBH) that are actively accreting matter, which generates an enormous amount of energy \citep{Rees84}. Among their many properties, active nuclei are characterised  by multi-scale outflows observed in various bands of the electromagnetic spectrum. \citep{Harrison14,Harrison18}. 

Ultra-fast outflows (UFOs) are high-velocity, massive and powerful winds of gas that are launched from the central regions of AGN. These outflows have velocities larger than $v_{out} \geq$ 10,000 km s$^{-1}$. They were detected in the X-ray spectra of AGN and were usually observed in the form of absorption lines of ionized gas  where the only dominant ions are helium-like and hydrogen-like Iron (e.g. \citet{Tombesi10}). Several studies have documented the presence of UFOs in specific AGNs \citep{Chartas09,Pounds14,Longinotti15,Krongold21} and according to systematic studies, outflows have been observed in 30-40$\%$ of nearby AGN (\citet{Matzeu23} and reference therein). \citet{Tombesi10} and \citet{Gofford13} suggest that UFOs are likely launched from the accretion disk scales. The mass outflow rate of these winds (0.01-1 M$_\odot$yr$^{-1}$) and their kinetic energy (10$^{42-45}$erg s$^{-1}$) are consistent with the predictions made by theoretical models of black hole feedback \citep{King10}.  

AGN feedback, which refers to the impact of AGN activity on the large scale environment, is thought to have a profound effect on the evolution of the host galaxy properties \citep{Hopkins10}. Simulations reproducing the galaxy luminosity function at high luminosities require a mechanism to regulate the star formation rate, and to produce the observed relationships between SMBH and their host galaxies \citep{DiMatteo05,Bower06}. AGN winds  have been  proposed as one of the mechanism behind Quasar feedback \citep{Kormendy13}. 
At present, the relation of the inner X-ray wind to galaxy-scale outflows, supports the idea that fast winds launched close to the AGN accretion disk may act as a trigger for altering star formation in the host galaxy and start feedback processes \citep{Feruglio15, Tombesi15, Longinotti18, Marasco20, Longinotti23}.  

NLS1 galaxies form a subclass of AGNs characterized by the presence of an H$\beta$ line with a FWHM $< 2000$ kms$^{-1}$ and an unusually prominent Fe II spectral lines \citep{Osterbrock85}. \citet{Brandt97}, \citet{Boller96}, showed that NLS1 exhibit X-ray properties different from the general population of Seyfert 1 galaxies. They have steep soft- and hard-X-ray spectra and show fast and large amplitude variability. These properties suggested that NLS1 have relatively smaller BH masses and higher Eddington accretion rates. As such NLS1 occupy one extreme end of the AGN "eigen-vector 1" (\citep{Boroson92}; see also \citet{Watson07}). It has been shown that NLS1 have super-solar metallicity \citep{Fields05, Fields055} and that they likely undergo secular growth of their BH (\citet{Grupe04,Mathur05,Mathur055}). Their host galaxies have pseudo-bulges, further supporting the interpretation of secular black holes growth \citep{Mathur12}. NLS1s may represent an early evolutionary stage of AGNs \citep{Mathur00}.

Recent studies have uncovered UFOs in soft X-ray X-ray high-resolution spectra of NLS1 \citep{Gupta13,Gupta15,Longinotti15,Parker17,Reeves19,Krongold21,Xu24}. High-resolution spectroscopy has revealed that soft X-ray UFOs can consist of multiple components with different velocities, ionization states, and column densities, and that, in some cases, they can be massive enough to produce feedback, as observed, for example, IRAS 17020 + 4544 \citet{Longinotti15,Longinotti23}).

Mrk 110 is a NLS1 at z=0.0353 with an X-ray luminosity  of L=1.0$\times$10$^{44}$erg s$^{-1}$ in the 2-10 keV band. Assuming a bolometric correction k=18 \citep{Duras20}, the bolometric luminosity for this source is L$_{BOL}$=1.85$\times$10$^{45}$erg s$^{-1}$. It was found variable in X-rays as well as in optical/UV wavelengths \citep{Grupe06,Kollatschny06}. \citet{Kollatschny03} estimated a SMBH mass of M$_{BH}$=1.8$\pm$0.4$\times$10$^{7}$M$_{\odot}$ via reverberation mapping methods. Given the Eddington luminosity of L$_{Edd}\sim$2$\times$10$^{45}$erg s$^{-1}$ (see 4.3) this gives a high accretion rate of $\lambda$/$\lambda_{Edd}\sim$0.82. 

As to its X-ray history, \citet{Boller07} reported the presence of broad and narrow emission lines in the {\it XMM-Newton} Reflection Grating Spectrometer (RGS) spectrum, finding that the distance from the SMBH to the emission line region ranges from 0.2 to 1 light day from the SMBH. \citet{Reeves21} reported in the RGS spectrum a broad component of the OVII emission line with a FWHM of 15900 $\pm$ 1800 km s$^{-1}$, that was modeled with a relativistic profile from gas in the inner accretion disc.  Based on  simultaneous observations of {\it XMM-Newton} and {\it NuSTAR}, \citet{Porquet21} showed that the dominant physical processes above 3 keV are hard Comptonisation in a hot corona and soft Comptonisation in a warm, optically thick corona. In this scenario, the warm corona accounts for the soft X-ray excess, while the hot corona produces the primary power-law continuum. According to these authors,  the latter with mildly relativistic disc reflection are responsible for the broad O VII and Fe K$\alpha$ emission lines. Regarding other observing bands, \citet{Panessa22} showed evidence for variability at 5 GHz on scales of days to weeks time-scales and suggested that emitting regions have a size smaller than $\sim$ 180 Schwarzschild radii. Considering  the high brightness temperature and the radio and X-ray variability, these authors proposed that synchrotron emission from a low-power jet and/or an outflowing corona is favored to explain the radio and X-ray timing properties.

More recently, \citet{Juranova24} report the presence of  broad X-ray lines in the grating spectrum that  can be accounted for by primarily UV-emitting gas in their model. \citet{Xiang25}, using {\it XRISM} observations of NGC 4151, revealed a stratified, multiphase accretion-disk wind in which warm absorbers, fast outflows, and UFO coexist, with mass outflow rates comparable to or exceeding the accretion rate. Notably, two UFO components carry sufficient kinetic power to produce significant AGN feedback, broadly consistent with magnetocentrifugal driving.
\citet{Mizumoto26}, using {\it XRISM} observations of PG 1211+143, resolved the Fe K absorption features into six distinct UFO velocity components ($-0.074$ to $-0.405c$), revealing a complex UFO forest structure. The inferred mass outflow rate ($\sim$~1 M$_{\odot}$ yr$^{-1}$) is comparable to the Eddington accretion rate, indicating that disk winds constitute a major channel of mass and energy release in near-Eddington AGN.

\section{Observations and Reduction}
\label{sec:obs}

\subsection{XMM-Newton} 
\label{XMM-Newton}

The {\it XMM-Newton} satellite observed Mrk 110 six times between November 2004 and April 2020 (see Table \ref{T1}). Five of these datasets are used in this study. The 2004 observation is not considered in this work due to the large temporal gap with respect to the other five observations, which were obtained within less than five months of each other. The data were processed using the \textit{rgsproc} task of the software System Analysis Software (SAS) 21.0. The spectra of the \textit{rgs1} and \textit{rgs2} cameras were combined with the RGSCOMBINE task. To improve the Signal-to-Noise Ratio (SNR S/N = counts/sqrt(counts)) and given the proximity in time of the observations and their nearly constant flux, the four spectra from 2019 and the one from 2020 were combined into one spectrum with the RGSCOMBINE task. The total exposure accumulated in this combined spectrum is  220.2 ks. The {\it EPIC-pn} exposures were processed by applying the standard filtering criteria through the SAS task {\it epproc}. The source spectra were extracted from circular regions of 40''. Background regions were selected in source-free portions of the CCD. Response matrices were provided by the SAS tools {\it arfgen} and {\it rmfgen}.

\subsection{NuSTAR} 
\label{NuSTAR}
{\it NuSTAR} \citep{nustar} observed Mrk 110 in January 2017, November 2019 and April 2020 (see Table \ref{T1}), with its two coaligned X-ray telescopes with corresponding Focal Plane Module A (FPMA) and B (FPMB). The total elapsed time of each of the three pointings exposures (Obs IDs 60201025002, 60502022002 and 60502022004) is 386 ks, 163 ks and 168 ks, respectively. The level 1 data products were processed with the {\it NuSTAR} Data Analysis Software (NuSTARDAS) package (v. 2.1.2). Cleaned event files (level 2 data products) were produced and calibrated using standard filtering criteria with the \textsc{nupipeline} task and the latest calibration files available in the {\it NuSTAR} calibration database (CALDB 20230321). Background spectra were extracted from source-free circular regions with a radius of 80 arcsec. Extraction radii for the spectra of the source were computed via an iterative process which leads to the maximization of the SNR in the 3--10 keV energy band, similar to the approach described in \citet{pico04}. We therefore adopted circular regions centered on the source with radii of 113 arcsec for the three FPMA spectra and 108 arcsec, 98 arcsec and 98 arcsec for the FPMB ones. FPMA spectra were binned in order not to over-sample the instrumental resolution more than a factor of 2.5 and to have a SNR greater than 5 in each spectral channel, the same energy binning was then applied to the FPMB spectra. The net observing times for the FPMA/FPMB spectra of the three data sets are 184.6 ks/184.3 ks, 86.8 ks/86.2 ks and 88.7 ks/87.9 ks, respectively.

\begin{table}[]
    \centering
    \caption{Log of the XMM-Newton RGS and NuSTAR observations of Mrk 110.}
    \label{T1}
    \footnotesize
    \begin{tabular}{ccccc}
    \hline
    \hline
         XMM-Newton     & Obs Date   & Exp  & Flux \\
         Obs ID         &            & ks   & $\times10^{-11}$erg cm$^{-2}$ s$^{-1}$ \\
                        &            &      & (RGS) 0.3-2.0 keV \\
    \hline
         0201130501 & 2004/11/03 & 47.4 & 3.22$\pm0.03$ \\
         0840220701 & 2019/11/03 & 43.6 & 2.06$\pm0.01$ \\
         0840220801 & 2019/11/05 & 43.0 & 2.61$\pm0.02$ \\
         0840220901 & 2019/11/07 & 40.6 & 2.73$\pm0.02$ \\
         0852590101 & 2019/11/17 & 44.5 & 2.76$\pm0.02$ \\
         0852590201 & 2020/04/06 & 48.5 & 2.13$\pm0.02$ \\
    \hline
         NuSTAR      &            &       & 3.0-15.0 keV  \\
    \hline
         60201025002 & 2017/01/23 & 184.5 & 4.46$\pm0.004$ \\
         60502022002 & 2019/11/16 & 86.7  & 2.87$\pm0.002$ \\
         60502022004 & 2020/04/05 & 86.7  & 3.34$\pm0.003$ \\
    \end{tabular}
\end{table}

\section{Spectral Analysis}
\label{sec:SA}
\subsection{RGS Data}

We carried out the spectral analysis of Mrk 110 RGS data using the unbinned spectrum, making use of the Cash statistics \citep{Cash79}, The software used for the spectral analysis was {\sc xspec} v.12.15.1 (X-Ray Spectral fitting Package). Fit parameters are quoted with 1$\sigma$ error bars throughout the paper.

\subsubsection{Modeling the Spectrum} 
\label{MS}

\begin{figure*}
\includegraphics[width=14.5cm]{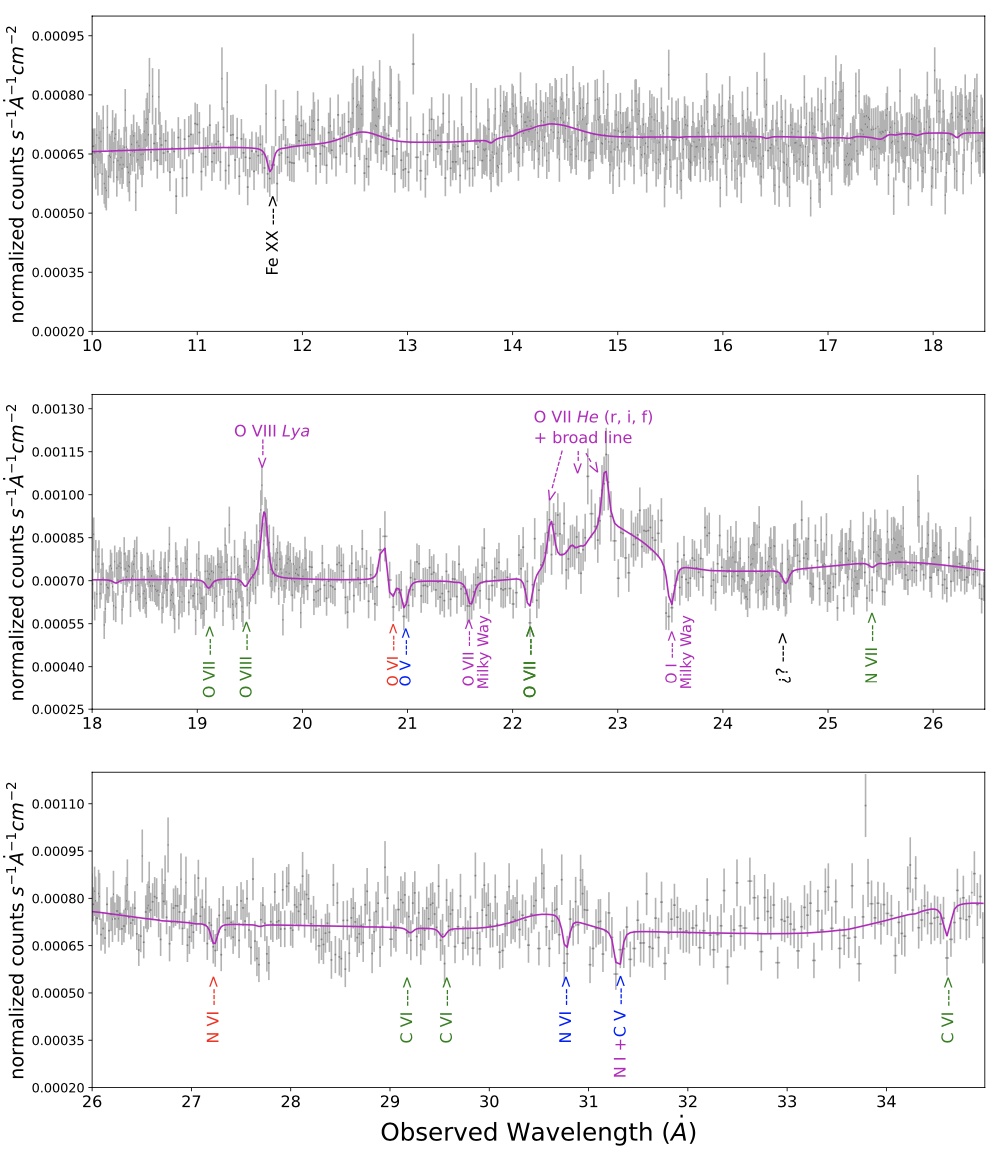}
\centering
\caption{RGS spectrum in the 10-35$~$\AA$~$range, fitted with a model including a power-law continuum and eighteen Gaussian components (four narrow and six broad in emission and eight in absorption). The O VIII Ly$\alpha$, O VII triplet ({\it r, i, f}) for narrow emission lines, the six broad emission and the absorption for O VII and O I of our Galaxy at z=0 are marked in magenta. The other absorption lines are reported in Table \ref{T2}. The feature at 31.3 is a blend of CV from UFO and and NI from CGM. The identifications were obtained after performing the spectral modeling and fitting of the data. The spectrum was binned by up to three instrumental channels for display purposes only and are shown in the observed wavelength. \label{fig:1}}
\centering
\end{figure*}

We fit the X-ray continuum in the 7-35$~$\AA$~$(0.35 - 2.5 keV) interval with a power law affected by Galactic absorption of column density $N_{H}=1.3\times10^{20}cm^{-2}$ \citep{Karlberla05} using the model TBabs \citep{Wilms00} of {\sc xspec}. The photon index of the power law is $\Gamma=2.21^{+0.05}_{-0.08}$, consistent with typical values obtained in the soft X-ray band for NLS1 \citep{Gallo18}. The model yields fit statistics of $C_{stat}$ = 3313 for 2797 degrees of freedom (d.o.f.). We found residuals hinting to the presence of emission and absorption features in the spectrum, therefore we included narrow components of the O VIII Ly$\alpha$ and O VII triplet ({\it r, i, f}) emission lines. We also included the absorption lines  by O VII at 21.601 associated with hot Galactic/CGM gas, and the NI absorption line at 31.3 $\dot{A}$ associated with cold Galactic gas (Figure \ref{fig:1}). The normalization of the NI absorption line is tied to that of the O I absorption line as such that their relative ionic column densities follow the ratio predicted by the TBabs component (i.e. $N_{NI}/N_{OI}=0.16).$ For $N_{H} = 1.3 \times10^{20}cm^{-2}$, this gives $N_{NI} \simeq 9.9 \times 10^{15} cm^{-2}$ and $N_{OI} \simeq 6.4 \times 10^{16} cm^{-2}$, hence $N_{NI}/N_{OI} \simeq 0.16$. We therefore fixed $Norm_{NI} = 0.16Norm_{OI}$, which yields $Norm_{NI} = -3.3\times 10^{-6}phcm^{-2}s^{-1}$.) The position and width of the narrow emission and absorption lines were fixed and we only left normalization free to vary. The model with the four narrow emission lines and the two absorption lines results in $C_{stat}$ = 3076/2792, and this model provides an improvement to the simple power law fit of $\Delta(C_{stat}/\mathrm{d.o.f})$ = 229/5.

Following the recent findings  by \citep{Juranova24} on the presence of broad emission lines in the RGS spectrum,  we added six broad emission lines as well as one Warm Absorber (WA) component. Although \citet{Juranova24} used a second WA component with log $\xi = 2.91$, the inclusion of this additional component in our model did not lead to any statistically significant improvement. Therefore, we opted to keep one single-component WA in our baseline model, fitted by the PHASE photoionization code (see details in  Section \ref{MUFO}). The model including six broad emission lines and one warm absorber (WA) provides a fit statistic of $C_{stat}$ = 2905/2783. The addition of the six broad lines plus the WA improves the fit by $\Delta$$C_{stat}$/d.o.f. = 171/9. Table \ref{T2} lists the best-fit parameters of the six broad emission lines, while Table \ref{T3} shows the parameters of the WA. The WA parameters, including their uncertainties, are consistent with those of the first WA reported by \citet{Juranova24}.

A visual inspection of the RGS spectrum fitted with this model still shows negative residuals throughout the entire spectral range.  These suggest absorption lines by ionized gas, so we performed a blind search for individual absorption lines by adding one narrow Gaussian component with negative intensity to our model, letting its position and intensity vary freely over the whole spectral range. The line width was fixed to 0.1$~$\AA$~$. This procedure was repeated for as many lines as required until the line significance drops below 2$\sigma$. Table \ref{T2} reports the 7 absorption lines detected above a 2$\sigma$ threshold with this procedure. The intensity of the line was divided by the positive 1$\sigma$ error to estimate the line significance. Finally we note that the line at $\sim$ 23.5 \AA ~is identified with O I from hot gas in the Galaxy. Figure \ref{fig:1} presents our best-fit model with all the lines reported in Table \ref{T2} plus the WA listed in Table \ref{T3}. 

Although most absorption lines in Table \ref{T2} seem to have marginal significance, two of them do show significance larger than 3$\sigma$. The presence of multiple negative residuals  suggest that they could be the result of a shallow absorber in the line of sight. However, the positions of the lines as reported in Table \ref{T2} do not align with any obvious  transitions in the reference framework of the source, suggesting the presence of an UFO, where absorption lines are significantly shifted from their expected positions. To test for the presence of  such outflowing gas, we proceeded to model the spectrum with a self-consistent ionized absorber, as described in the following.

\begin{table*}[h]
    \centering
    \caption{List of Emission and Absorption Lines Detected in the RGS data.}
    \label{T2}
    \scalebox{0.9}{
    \begin{tabular}{ c c c c c c c }
    \hline
    \hline
        Obs. $\lambda$          & Rest Wavelengths & Intensity                      & Sign     & Line       & Component  \\ 
        $\dot{A}$               & z = z(Mrk110)    & $10^{-5}$ph cm$^{-2}$ s$^{-1}$ & $\sigma$ & ID         &            \\ 
    \hline
        Emission                &                  &                                &          &            &            \\
        12.14                   &                  & 1.48$^{+0.65}_{-0.64}$         & 2.3      & Ne X       & Broad      \\
        13.85                   &                  & 2.95$^{+0.58}_{-0.57}$         & 2.5      & Ne IX      & Broad      \\
        18.96                   &                  & 2.56$^{+0.34}_{-0.33}$         & 7.5      & O VIII     & Narrow     \\
        21.6                    &                  & 2.79$^{+0.56}_{-0.57}$         & 4.9      & O VII      & Narrow     \\
        21.8                    &                  & 2.13$^{+0.58}_{-0.57}$         & 3.7      & O VII      & Narrow     \\
        22.101                  &                  & 6.75$^{+0.82}_{-0.8}$          & 8.4      & O VII      & Narrow     \\ 
        22.101                  &                  & 18.52$^{+2.24}_{-2.2}$         & 8.4      & O VII      & Broad      \\
        24.78                   &                  & 7.33$\pm 1.4$                  & 5.2      & N VII      & Broad      \\
        29.53                   &                  & 3.32$\pm 1.4$                  & 2.4      & N VI       & Broad      \\
        33.74                   &                  & 3.07$\pm 0.4$                  & 7.5      & C VI       & Broad      \\
    \hline
        Absorption              &                  &                                &          &            &            \\
        11.69$\pm0.02$          & 12.86            & -0.82$^{+0.23}_{-0.32}$        & 3.6      & Fe XX      & ?          \\
        20.84$^{+0.06}_{-0.05}$ & 22.02            & -0.66$^{+0.32}_{-0.7}$         & 2.0      & O VI       & UFO1       \\
        20.97$\pm0.01$          & 22.37            & -1.17$^{+0.29}_{-0.6}$         & 4.0      & O V        & UFO2       \\
        24.6$^{+0.01}_{-0.07}$  &                  & -0.55$^{+0.26}_{-0.41}$        & 2.1      & ?          & ?          \\
        27.23$^{+0.02}_{-0.12}$ & 28.87            & -0.82$^{+0.41}_{-0.29}$        & 2.0      & N VI       & UFO1       \\
        30.77$\pm0.01$          & 32.75            & -1.39$^{+0.51}_{-0.57}$        & 2.7      & N VI       & UFO2       \\
        31.3$^{+0.06}_{-0.01}$  & 33.42            & -1.49$^{+0.68}_{-0.53}$        & 2.7      & C V        & UFO2       \\
    \hline
                                &                  &                                &          & CGM        &            \\
        21.602                  &                  & -1.37$\pm 0.43$                & 3.1      & O VII      &            \\
        31.3                    &                  & -0.33                 & 3.9      & N I        &            \\
    \hline
    \multicolumn{7}{ c }{\textbf{Note.} The identification of lines were made by matching this phenomenological fit to the} \\
    \multicolumn{7}{ c }{physical model in Table 3.}
    \end{tabular}}
\end{table*}

\subsubsection{Modeling the Ultra Fast Outflow} \label{MUFO}

To achieve a physical description of the absorption in the RGS spectrum of Mrk 110, we use a self-consistent model, specifically the photo-ionization code {\sc phase} \citep{Krongold21, Longinotti15}. The model assumes a slab of ionized gas in a plane-parallel geometry, and it is described by four parameters: the ionization parameter\footnote{The dimensionless ionization parameter is defined as $U = \frac{Q}{4\pi R^2cn_e}$ where $Q$ is the rate at which the source emits ionizing photons (i.e., photons per second), $c$ is the speed of light, and $n_e$ and R are the electron density and the distance of the gas from the X-ray source, respectively.}, the equivalent hydrogen column density of the slab $N_H$, the outflow velocity of the wind $v_{out}$ and the micro-turbulence velocity of the medium. Given that the Gaussian absorption lines in the spectrum are narrow, we fix the micro-turbulence velocity to 10 $km~s^{-1}$.

\subsubsection{The Significance of the UFO Components} \label{AIC}

While fitting the data with the PHASE components, we evaluated the robustness of a model including a UFO using two different statistical methods.

First, we used the Akaike Information Criterion values \citep{Akaike74}, as done by \citet{Hebbar19}. These values are defined as: 

\begin{equation}
    AIC = 2k + C_{stat}
\end{equation}

where $k$ is the number of parameters and $C_{stat}$ is the value of the $C-$statistic for the best fit with a given model\footnote{We note that, since we are using the unbinned RGS data, we have a large number of d.o.f. in our models, and we do not require the correction of the Akaike values for small sample sizes.}. The likelihood that the model including the UFO component is statistically preferred over the model without it, is given by the quantity exp((AIC(UFO) - AIC(No-UFO))/2). \citet{Hebbar19} pointed out that the inverse of this quantity represents the factor by which the more complex model, which in our case contains the UFO, is preferred over the simpler model.

The second method to estimate the statistical significance of the UFO components consisted of Monte Carlo simulations. To accomplish this, we followed the procedure in \citet{Longinotti15} (hereafter L15). We assumed the phenomenological continuum model described in Section \ref{MS} (comprising a power law with Galactic absorption plus 10 emission lines and one Galactic absorption line) as our baseline model for the simulations. Then, we incorporated each  identified UFO components into the baseline model, as done in the actual spectral fitting. For instance, in the simulations, we included only the baseline model to evaluate the significance of UFO1 and baseline+UFO1 to assess the significance of the UFO2. For each UFO component, we generated 1000 synthetic spectra using the {\sc xspec} package, each with the same photon statistics as the real data set. We compared the distributions of $\Delta C_{sim}$ between the model with and without the addition of the UFO component to the values measured in reality ($\Delta C_{true}$). The fraction of simulations where $\Delta C_{sim}$ is lower than $\Delta C_{true}$ tells us the confidence level, and this conclusion led to the significance values reported in the last column of Table \ref{T3}.

\subsubsection{Fitting the Ultra Fast Outflow} \label{FUFOs}

\begin{figure*}
\includegraphics[width=14.5cm]{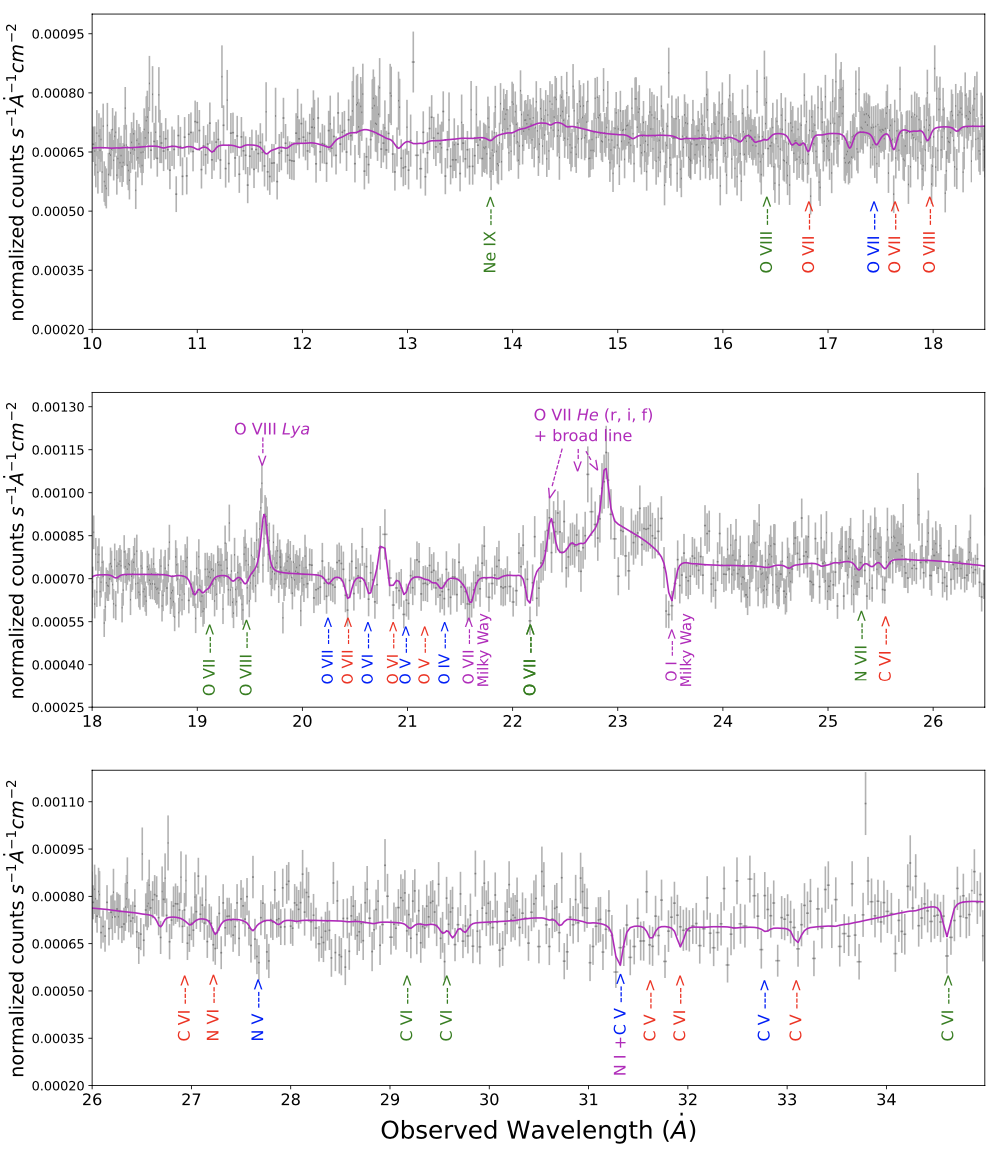}
\centering
\caption{RGS spectrum in the 10-35 $\dot{A}$ range, fitted with a model including a power-law continuum attenuated by one WA and 2 UFO components. The absorption lines produced by the WA are labeled in green, for UFO1 are labeled in red and those by UFO2 are marked in blue. The O VII and O I from the CGM of our Galaxy at z=0 are marked in magenta. The spectrum was binned by up to three instrumental channels for display purposes only and are shown in the observed wavelength. Absorption lines marked in green are associated with the WA component, while those shown in red and blue correspond to UFO1 and UFO2, respectively. These identifications were obtained after performing the spectral modeling and fitting of the data. \label{fig:2}}
\centering
\end{figure*}

Initially, we fit the data with a model consisting of a power law continuum absorbed by a slab of photo-ionized gas modeled by a {\sc phase} component in the 7-35$~$\AA$~$(0.35 - 2.5 keV) range, plus Galactic absorption, plus the broad and narrow emission lines, plus the CGM absorption line and the WA described in Section \ref{MS}. The ionization parameter, the column density, and the outflow velocity of the absorber were left free to vary. The addition of one PHASE component improves the fit of $\Delta$C = 21/3 d.o.f. ($C_{stat}$ = 2884 for 2780 degrees of freedom), indicating the likely presence of ionized gas  along the line of sight. The absorber (hereafter UFO1) has an outflow velocity of $v_{out}\approx 0.053c$ (see Table \ref{T3}). Given its low level of ionization (Log $U \approx -1.43$) and column density (Log $N_H=19.7$), UFO1 produces absorption by ions such as O VIII, O VII, O VI, O V, C VI, C V, C IV and N VI (red labels in Figure \ref{fig:2}). The significance of UFO1 was evaluated as described in Section \ref{AIC}. Comparing AIC statistics suggests that the model including UFO1 is preferred by a factor of 1808 over the model that only considers the continuum. Using Monte Carlo simulations UFO1 have a statistical significance of $\sim4\sigma$.

After the addition of UFO1, other residuals remain in the spectrum. Therefore, we continued fitting the data by iteratively adding additional PHASE component to the model (as we did with UFO1, for each component, the ionization parameter, column density, and outflow velocity were free parameters). One additional statistically significant PHASE component was found. This component is hereafter designated as UFO2. Table \ref{T3} presents the results of our best fit and Figure \ref{fig:2} shows the best fit of the RGS spectrum.

UFO2 has a slightly higher velocity than UFO1, its value is $v_{out}\approx0.062c$. UFO2 has low ionization (Log $U\approx-1.82$) and column density of (Log $N_H\approx19.7$). The absorption lines associated to UFO2 are O VII, O VI, O V, O IV, N V and C V (blue labels in Figure \ref{fig:2}). UFO2 improves the fit, with $\Delta$ = 18/3 d.o.f. ($C_{stat}$ = 2866). The AIC statistics indicates that UFO2 improve the fit by factor of $\approx$ 403 and the Monte Carlo simulations from UFO2 shows a statistical significance of $\sim3\sigma$. 

The final best-fit model includes the power-law continuum, plus four narrow and 6 broad emission lines, plus the CGM absorption line, plus one WA and the two UFOs components reported in  Table \ref{T3}. The final RGS best fit is shown in Figure \ref{fig:2}.

Given that the absorption lines imprinted by these absorbers are shallow and difficult to recognize, we carried out further checks by allowing identification by other species. First,  we tested if the O V line in UFO2 could be instead  O IV or O VI in two different ways, i.e  by changing the ionization parameter and changing the outflow velocity of the PHASE component. In both cases, the statistic got worse with the case of the line being  O IV yielding an  improvement of  $\Delta$ = 5/3 d.o.f. ($C_{stat}$ = 2900); in the case of the line being identified as O VI, the improvement in the statistics was even smaller $\Delta$ = 2/3 d.o.f. ($C_{stat}$ = 2903). We tried the same test for UFO1 where  O VII might be instead identified as O VI or O V.  In both cases the model with the alternative identifications did not improved the statistic with respect to our proposed interpretation. In addition the  absorption pattern corresponding to the alternative identifications result in a model that produces other absorption features clearly not consistent with the data. With this test we further corroborate that the identification of the lines is correct and the best fit reported Table \ref{T3} is the most consistent with the data. 

Our fit still shows some negative residuals, particularly around 11.69 and 24.6 $\dot{A}$ (see Table \ref{T2}), where negative Gaussian profiles with significances of 3.6 and 2.1 $\sigma$ were required. The line at 11.69 is consistent with absorption by Fe XX. The feature at 24.6 $\dot{A}$ cannot be accounted for by adding another significant UFO component or by the presence of a Galactic line at z=0, and therefore, it remains unidentified in our analysis. The line at 30.77 $\dot{A}$ could be consistent with absorption by C V at the same velocity of UFO2, however, this component does not have enough neutral material to produce this feature.

As a final remark we noted the spectrum shows positive residuals around 23 to 23.5 $\dot{A}$. These residuals were modeled by \citet{Reeves21} with a relativistic line profile. Since they fall right next to the galactic line O I (23.5 $\dot{A}$) and none of the detected absorption features fall in this spectral portion, we decided not to include the modeling of these residuals in our analysis.

\begin{table*}[]
    \centering
    \caption{List of WA and Ultra-fast Wind Detected in the RGS and NuSTAR data of Mrk 110.}
    \label{T3}
    \scalebox{0.85}{
    \begin{tabular}{ccccc|cc}
    \hline
    \hline
         Component      & Log U                   & Log$N_H$                & $v_{out}$             & Statistics                & AIC  & Factor \\
         instrument     &                         & (cm$^{-2}$)             & (km s$^{-1}$)         & $\Delta$ C/$\Delta$ $\nu$ & 3114 &        \\
    \hline
         WA             & -1.11$^{+0.15}_{-0.11}$ & 19.43$^{+0.1}_{-0.16}$  & 2900$\pm100$          & 16/3                      & 2943 & 148    \\
         UFO1           & -1.43$\pm0.12$          & 19.74$^{+0.08}_{-0.14}$ & 26600$^{+200}_{-300}$ & 21/3                      & 2928 & 1808   \\
         UFO2           & -1.82$^{+0.12}_{-0.16}$ & 19.7$\pm0.11$           & 29300$^{+110}_{-350}$ & 16/3                      & 2916 & 403    \\
    \hline
                        &                         &                         &                       & $\chi^2/d.o.f.$           & 528  &        \\
         UFO3  (NuSTAR) & 2.19$^{+0.17}_{-0.1}$   & 22.81$^{+0.36}_{-0.21}$ & 64300$\pm900$         & 18/3                      & 512  & 2981   \\
    \hline
    \end{tabular}
    }
\end{table*}

\subsection{Searching for the highly ionized fast wind in CCD spectra } \label{NUSTAR}

Motivated by the findings of the ultra-fast wind in the RGS data, we analysed the NuSTAR and XMM-Newton EPIC pn spectra to search for higher ionization absorption by Fe~XXV or/and Fe~XXVI. 

\subsubsection{The NuSTAR spectrum}
We begin our analysis with NuSTAR since it has a broader bandwith than the XMM-pn, so that we can later compare it with the XMM-Newton pn data. We used $\chi^2$ statistics, given the binning applied to the data. First, we simultaneously fit the NuSTAR spectra from the three observations by adding an energy-independent factor (model constant of XSPEC) to the continuum for small calibration uncertainties from the FPMA and FPMB detectors. 

We fit the X-ray continuum in the interval of 3-15 keV, using a power law modified by Galactic absorption of column density of  $N_H$=1.3$\times$10$^{20}$cm$^{-2}$ \citep{Karlberla05} with the model TBabs \citep{Wilms00} of XSPEC. The photon index of the power law is $\Gamma =1.78 \pm 0.01$. This model provides an acceptable representation of the data, with a $\chi^2/d.o.f. = 795/435$. However, we visually noted evidence of residuals pointing to the presence of an emission line from Fe I-XII K$\alpha$ fluorescence at 6.33$\pm0.02$ keV. Adding a Gaussian profile to fit this line component improves the fit statistic to  $\chi^2/d.o.f. = 528/433$. 

\begin{figure}
\includegraphics[width=\columnwidth]{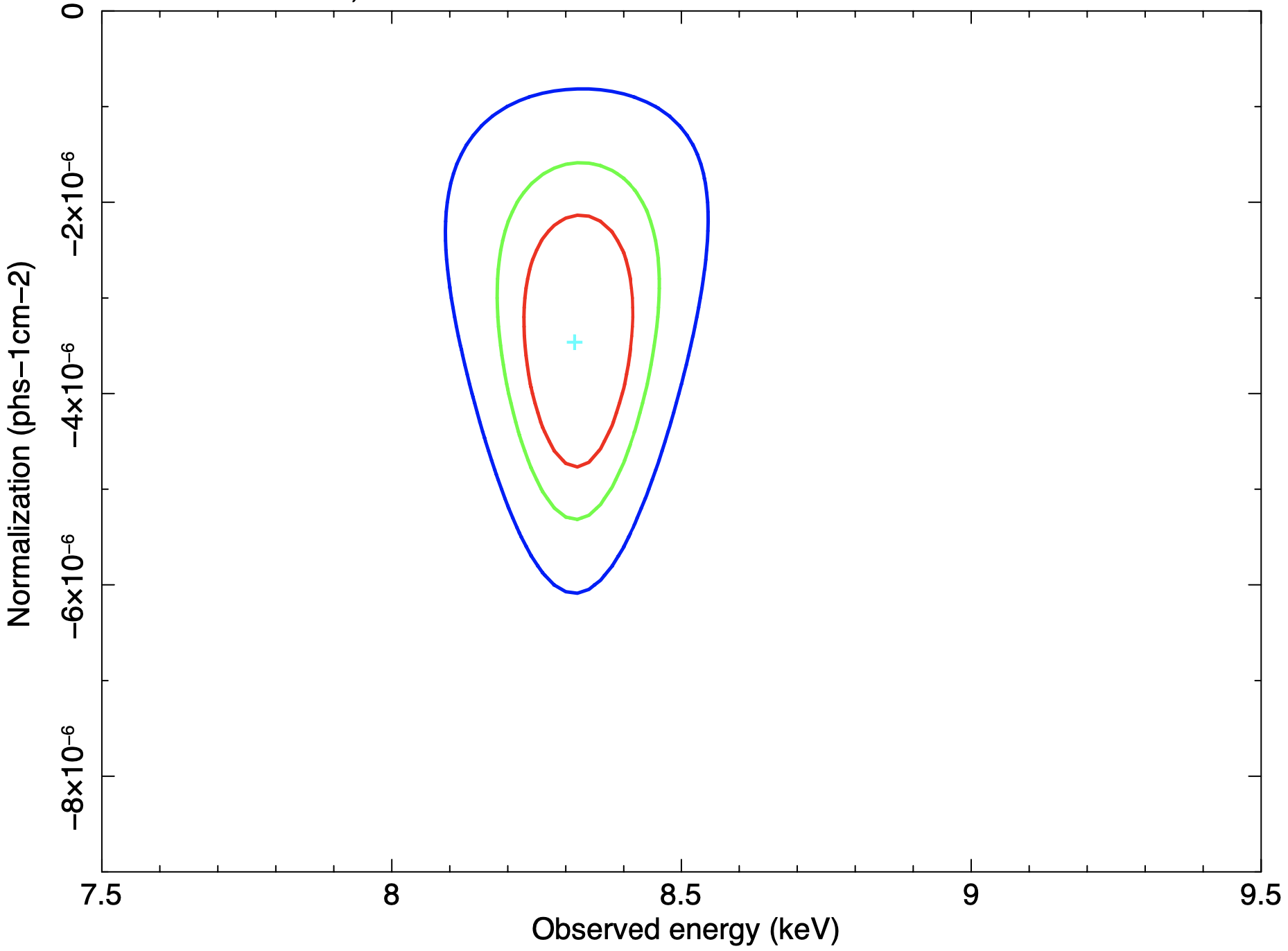}
\caption{Confidence contour plot between the absorption line centroid energy and flux (68\% (blue), 90\% (green), and 99\% (red)), from the simultaneous fit for the three observations of NuSTAR data. \label{fig:3}}
\end{figure}

In addition to the  Fe I-XII K$\alpha$ fluorescence line, we found evidence for negative residuals in this spectral band. These negative residuals were fit by adding a narrow Gaussian line to the above model with the intensity allowed to vary between positive and negative normalization. The position of the line was also free to vary. The statistical improvement derived by the addition of the absorption feature was $\Delta \chi^2 = 16/2$. The position of the line is 8.31$^{+0.06}_{-0.09}$ keV where absorption features by Fe~XXV or Fe~XXVI from highly ionized gas at mildly relativistic velocities might be expected. Figure \ref{fig:3} shows the 68$\%$, 90$\%$, and 99$\%$ confidence contours between the absorption line energy centroid vs the line intensity, confirming  a statistical significance larger than 3~$\sigma$. Assuming that absorption is due to Fe~XXV or Fe~XXVI, the position of the line would indicate the presence of absorbing gas with a velocity ranging from 64,400$^{+3,300}_{-5,300}$ to 85,500$^{+3,700}_{-5,500}$, respectively.

\begin{figure*}
\includegraphics[width=15cm]{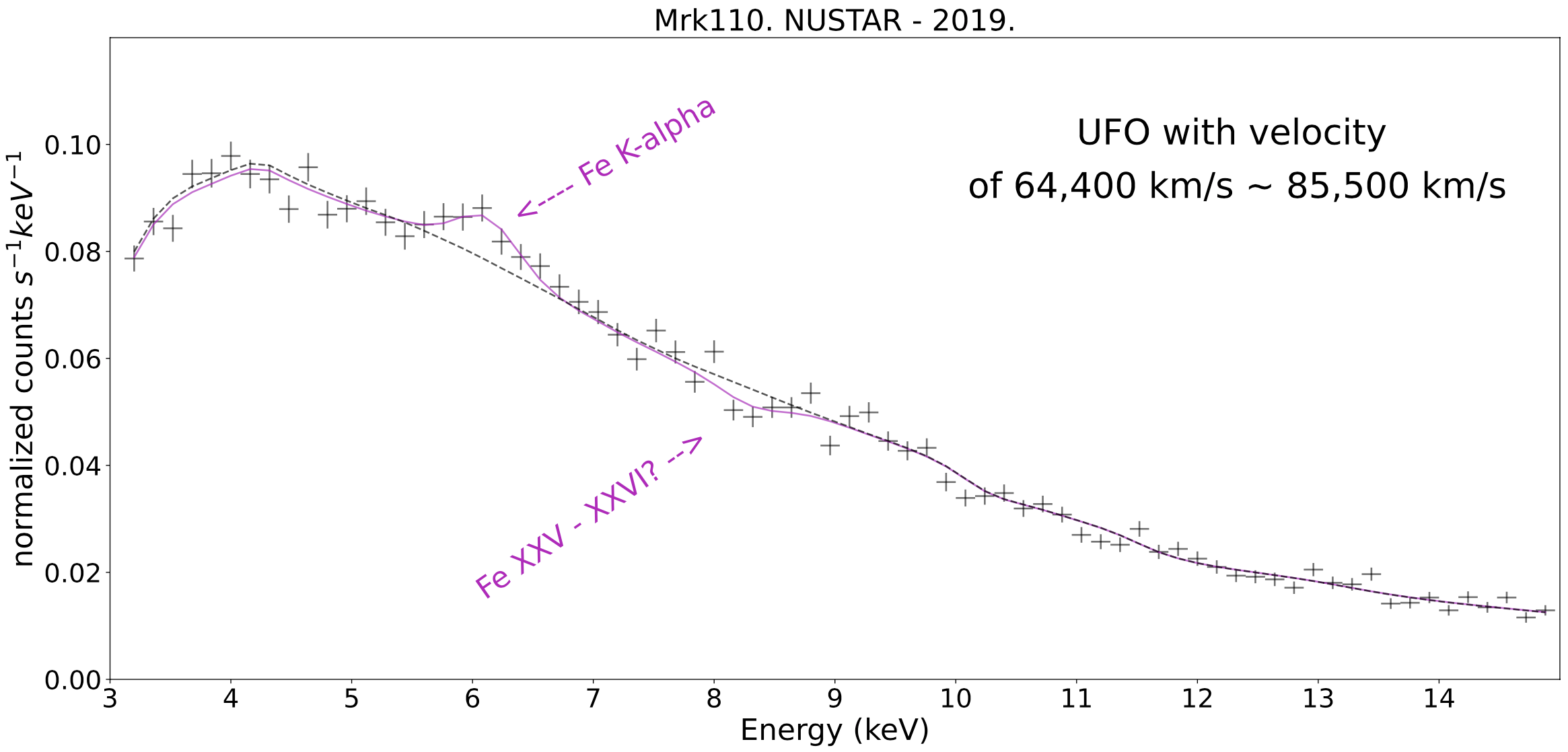}
\centering
\caption{\textbf{NuSTAR spectra of Mrk 110 in the 3-15 keV range, fitted with a model including a power-law continuum attenuated by Galactic absorption. One emission line produced by Fe I-XII K$\alpha$ fluorescence  and one absorption line interpreted as outflowing gas from  Fe XXV or Fe XXVI are included in the model. \label{fig:4}}}
\centering
\end{figure*}

After carrying out the simultaneous analysis on the 3 NuSTAR observations we decided to look for the line in each observation separately. In the year 2017, the line was measured with a statistical significance of 2$\sigma$ , in 2019 the significance reached a level above 3$\sigma$, while in 2020, a significance of 2$\sigma$ was observed. Figure \ref{fig:4} shows the model fitted to the NuSTAR data of  2019.

\subsubsection{The joint NuSTAR-XMM-Newton EPIC-pn spectrum}

We decided to support the analysis of NuSTAR spectrum by including in our analysis the EPIC-pn spectra from 2019 and 2020, therefore the ones contemporanous to NuSTAR (see Table 1). We used the combined pn spectrum for these two epochs and produce a joint fit NuSTAR-XMM. The continuum baseline model of NuSTAR, consisting of a power-law and a Fe I-XII K$\alpha$ fluorescence at 6.33$\pm0.02$ keV line, was applied to the joint CCD spectrum. The inclusion of the XMM-Newton pn spectra allowed us to find an additional emission feature at  6.67$^{+0.03}_{-0.04}$ keV, resulting in a significant improvement of $\Delta \chi^2 = 12/2$ for 2 d.o.f. This line is consistent with emission from FeXXV, as commonly found in several AGN e.g. \citet{Longinotti07} \citet{Bianchi08} and \citet{XRI1365}.  We also attempted to fit additional absorption lines at 6.97 keV and 7.06 keV respectively corresponding to FeXXVI and Fe I K$\beta$ transitions; however, these features led to only marginal improvements ($\Delta \chi^2 < 2$), therefore they were not included in the final model. Once the two significant emission lines were included in the model, we search for the absorption line in the joint fit and found it at 8.33$^{+0.07}_{-0.06}$ keV, resulting in an improvement of  $\Delta \chi^2 = 17/2$. The EPIC-pn spectrum does not show the absorption feature as clearly as the NuSTAR spectrum when inspected alone. However, in the join EPIC-pn/NuSTAR fit, the absorption line is recovered at $E=8.33^{+0.07}_{-0.06}$ keV, with an improvement of $\Delta \chi^2 = 17$ for two additional free parameters. This centroid energy is fully consistent with the value obtained from the NuSTAR-only fit, $E=8.31^{+0.06}_{-0.09}$ keV. We therefore conclude that the EPIC-pn data are consistent with the presence of the absorption feature, although they do not independently provide a stronger detection than NuSTAR.

\subsubsection{Applying a physical model to NuSTAR spectrum }
After establishing that the absorption feature is recovered in the joint EPIC-pn/NuSTAR fit with a centroid energy consistent with the NuSTAR-only analysis, we proceeded as in Section 3.1.4 and applied a PHASE component to the NuSTAR data. Since the joint EPIC-pn/NuSTAR fit does not provide significantly different line parameters with respect to NuSTAR alone, and since the EPIC-pn spectrum does not add strong independent constraints on the absorber, we performed the physical modeling and the computationally expensive Monte Carlo simulations using the NuSTAR data only. Therefore, the significance quoted below for UFO3 refers to the NuSTAR-based PHASE analysis, while the EPIC-pn spectrum is used as an independent consistency check. Upon multiple iterations with distinct parameter sets and comparison to the transitions predicted by PHASE, we could determine that the absorption line is most likely identified as Fe XXVI.  We, therefore, labeled this additional absorber UFO3. The associated parameter values for UFO3 are Log $U \approx 2.2$, Log $N_{H} \approx 10^{23}cm^{-2}$ and $v_{out} \approx 0.214c$. The model improved the fit by $\chi^{2}/d.o.f. = 18/3$. Given the high ionization state of UFO3, the expected absorption lines in the spectrum are Si XIV, S XVI, Fe XXIV, Fe XXV, and Fe XXVI.  Unfortunately, none of these lines can be seen in the RGS data (they all fall below 10~$\dot{A}$), therefore a cross-check for this particular UFO component cannot be properly performed. Finally, the AIC statistics indicates that the inclusion of the UFO detected in the NuSTAR spectrum improves the fit by a factor of $\approx$ 2981. To assess its statistical significance, we performed Monte Carlo simulations following the procedure described in Section 3.1.4. We generated 1000 synthetic spectra based on a simultaneous fit to the 2017, 2019, and 2020 observations, adopting a baseline model consisting of a power-law plus two Gaussian lines at 6.33 keV and 6.67 keV. From this analysis, we obtained a statistical significance of $\sim$ 4$\sigma$ for the UFO component.

\section{Discussion}

Despite the moderate significance of the absorbers found in our analysis, two UFO systems are detected at $\gtrsim4\sigma$. Previous studies  by \citet{Reeves21} and \citet{Porquet21} were primarily focused on emission originating from the disk and the warm and hot corona. This may explain why no absorption was reported. In the following, we discuss our interpretation  of the shallow ultra fast outflow observed in Mrk 110.

\subsection{Interplay of Light and Heavy UFOs}

Mrk 110 is another case of a Narrow Line Seyfert 1 Galaxy with multiple UFO components. Here, three different UFO systems were detected. Two of these outflows (UFO1 and UFO2) have quite similar column densities (Log $N_H \sim 19.7$ cm$^{-2}$, see Table 3) and show the lower range  of outflow velocity  ($\sim$26,000-29,000 km/s) measured in this analysis. We will refer to these systems as "light" UFOs. Within the errors, UFO1 and UFO2 have remarkably similar ionization parameters.  The third UFO (UFO3) reported in NuSTAR spectra, is clearly different from the former two systems, with 2 to 3 orders of magnitude larger column density and ionization state. UFO3 is also the fastest, with an outflowing velocity larger by a factor of 2 than the light UFOs. We will refer to this system as a "heavy" UFO.

Mrk 110 exhibits different features compared to a simple wind produced by one or two lines of highly ionized Fe. This is in line with other recent studies where stratification in the flow was observed in NLSy1 galaxies \citep{Longinotti15,Krongold21}.  As in those objects, light UFOs in this source are shallow and moderately ionized. One  current view of stratified light UFOs is that they are associated with the shock of a heavy wind with the interstellar medium (ISM) of the host galaxy producing high-velocity entrained clumps of gas, and/or with related hydrodynamical instabilities in the flow \citep{Pounds11, Sanfrutos18}. The complexity found in Mrk 110 with the different velocity in the heavy component poses yet additional clues to this view. The presence of the heavy UFO  in other sources has been interpreted in terms of perpendicular motion of the gas with respect to our line of sight \citep{Pinto17, Reeves18}. The particular view offered by Mrk 110 suggests that both processes (shocks and perpendicular motions) are at play, with a scenario more complex than the simple radial flows explored in numerical and analytical models of BH-ISM interaction and feedback \citep{King10}. An overall scenario seems to arise, where the wind consists of a multi-phase heavy flow interacting with the host galaxy ISM, resulting in the light components.

\subsection{Narrow Line Seyfert 1 and its Relationship with Ultra-Fast Outflows}

The detection of high velocity winds in Mrk 110, a NLS1 galaxy, adds to the growing number of sub-relativistic winds found in this type of AGN. The high accretion rates in NLS1 may play a role in the launching mechanism of these winds, which have been suggested to be radiatively driven \citep{Matzeu17, Pinto18}. We note however that recent research has questioned the efficiency of this process in accelerating gas to subrelativistic speeds \citep{Luminari21}. Nevertheless, whatever the process responsible for launching the wind, a tight connection between high accretion and UFOs is clearly emerging, given the steady increase in the number of detections of ultra-fast outflows in NLS1 over the last decade \citep{Gupta15, Longinotti15, Parker17, Kosec18, Reeves19, Krongold21}. 

Another tantalizing piece of the puzzle relates to the UFO and radio emission connection, specifically in highly accreting AGN with fast X-ray outflows, where pc-scale radio jets have been discovered (see \citet{Longinotti23} and reference therein). Indeed, a recent  VLBA-X-ray campaign on Mrk 110 \citep{Panessa22} revealed a variability pattern that among other alternatives, may be interpreted as due to a jetted or outflowing structure on sub-pc to pc scales. The present discovery of UFO processes in this source opens the door for further investigations on the coincidence of low power radio outflows/jets and X-ray winds in AGN.

\subsection{Feedback implications} \label{feedback}

Using simple energetic considerations, we can estimate if UFOs have the capacity to carry enough energy to suppress star formation in their host galaxy and influence its evolution. In accordance with \citet{Hopkins10}, a wind that transports energy outward in an energy-driven mode requires a minimum power level for feedback equivalent to 0.5$\%$ of the Eddington luminosity. As we illustrated below, UFO3 observed in Mrk 110 align with this criterion, suggesting their potential to satisfy this condition. 

To estimate the energetics of the outflows in Mrk 110, we assume that the outflow velocity is larger than or equal to the escape velocity at the launch radius. We take this to be $r=2GM/v^{2}_{out}$. Under this condition, the mass outflow rate for a radial wind in spherical geometry crossing our line of sight can be estimated by 

\begin{equation}
    \dot{M}_{out} = 4 \pi \mu m_p r N_H v_{out} C_f 
\end{equation}

where $m_p$ and $\mu$ correspond to the mass of the proton and to the mean atomic mass per particle ($\mu$=1.4), $C_f$ to the covering fraction of the outflow, and $N_H$ and $v_{out}$ to the column density and the observed velocity of the outflow, respectively \citep{Tombesi15}. Following \citet{Longinotti15}, we can express this formula in units of $10^{21}$cm$^{-2}(N_H=N_{21}\times10^{21}$cm$^{-2})$ and 10,000 km s$^{1}$ ($v_{out}=v_{10}\times10^4$km s$^{-1}$) as 

\begin{equation}
   \dot{M}_{out} = 7.8 \times 10^{15} \left(\frac{M}{M_{\odot}}\right) \left(\frac{N_{21}}{v_{10}}\right) C_f~g~s^{-1}
\end{equation}

\begin{equation}
   \dot{M}_{out} = 1.24 \times10^{-10} \left(\frac{M}{M_{\odot}}\right) \left(\frac{N_{21}}{v_{10}}\right) C_f~M_{\odot}~yr^{-1}.
\end{equation}

The energy outflow rate, or kinetic luminosity, is 

\begin{equation}
    \dot{E} = \frac{1}{2} \dot{M}_{out}v^2_{out}
\end{equation}

\begin{equation}
    \dot{E} = 3.9 \times10^{33} \left(\frac{M}{M_{\odot}}\right) N_{21} v_{10}~C_f~erg~s^{-1}. 
\end{equation}

We found that UFO3 is the only component that could produce feedback. Using M = 1.8$\times10^7M_{\odot}$, L$_{Edd}$ = 1.26$\times10^{38}(M/M_{\odot})$ = 2.2$\times10^{45}$ergs$^{-1}$, $N_H$ and $V_{out}$ from Table \ref{T3} for UFO3 yields 

\begin{equation}
    \dot{M}_{out} \approx 2.24 \times10^{-2} C_f~M_{\odot}~yr^{-1}
\end{equation}

and

\begin{equation}
    \dot{E} \approx 2.91 \times10^{43} C_f~erg~s^{-1}.
\end{equation}

The \citet{Hopkins10} criterion for triggering feedback by an energy-conserving wind for UFO3 is 

\begin{equation}
    \frac{\dot{E}}{L_{Edd}} = 0.0129C_f > 0.005
\end{equation}

for very conservative covering factors. If C$_f$ for UFO3 is $C_f >$ 0.39 this component is capable of significant feedback over their host galaxy. It is important to highlight that these estimates come with certain caveats related to the geometry of wind, which contributes to the computation of $\dot{E}$. Another source of uncertanty is the black hole mass determination, which goes into the estimation of $L_{Edd}$. Considering this, the above ratio should be interpreted with caution. 

Previous studies have shown that 30$\%$ to 40$\%$ of Seyfert galaxies show UFOs \citep{Tombesi10, Gofford13}. If we consider this range of 30–40$\%$ as a proxy for the wind covering factor, the heavy UFO in Mrk 110 may satisfy the theoretical feedback condition for covering factors near the upper end of this range. An evolving perspective continues to grow: Winds in NLSy1 galaxies appear to have full potential for substantial AGN feedback in their host galaxies.

\section{Conclusions}

The presence of the UFOs shown herein is based on the simultaneous detection of three different UFOs components in {\it XMM-Newton} and {\it NuSTAR}. These UFOs are fitted  with {\sc phase} a self-consistent absorption model, and despite their moderate significance, their properties have been fully characterized and validated through Montecarlo methods, showing a combination of shallow absorbers with heavier components. 

We show that under certain physical assumptions, the heavy UFO is capable of carrying sufficient energy to induce feedback in the host galaxy. The overall wind structure in this source, with heavy and light UFOs, represents a new example that ultra fast winds with a stratified structure are commonplace in Narrow Line Seyfert 1 galaxies, therefore providing a new candidate where the UFO may arise from a shock of the wind with the ISM.

\section{Acknowledgment}

The authors of this work thank the anonymous referee for his/her valuable comments. We thank the Instituto de Astronom\'ia at Universidad Nacional Aut\'onoma de M\'exico (IAUNAM). ALL, OSM and CV-C acknowledge support from DGAPA-PAPIIT IA103625. Lara-DI acknowledges support from Secretaria de Ciencia, Humanidades, Tecnolog\'ia e Innovaci\'on (SECIHTI) through the Estancias Posdoctorales por M\'exico Convocatoria 2025 (Estancia Posdoctoral Iniciales) grant program. M.G. acknowledges financial support from CONAHCYT through the SNII Research Assistant programme, granted to support the research activities of A.L.L., a Level III National Researcher. MSL acknowledges support from PID2022-136598NB-C33 by MCIN/AEI/10.13039/501100011033.


\renewcommand{\refname}{REFERENCES}
\bibliographystyle{rmaa}
\bibliography{RMxAA}

@ARTICLE{Akaike74,
       author = {{Akaike}, H.},
        title = "{A New Look at the Statistical Model Identification}",
      journal = {IEEE Transactions on Automatic Control},
         year = 1974,
        month = jan,
       volume = {19},
        pages = {716-723},
       adsurl = {https://ui.adsabs.harvard.edu/abs/1974ITAC...19..716A}
}

@ARTICLE{Bianchi08,
       author = {{Bianchi}, Stefano and {La Franca}, Fabio and {Matt}, Giorgio and {Guainazzi}, Matteo and {Jimenez Bail{\'o}n}, Elena and {Longinotti}, Anna Lia and {Nicastro}, Fabrizio and {Pentericci}, Laura},
        title = "{A broad-line region origin for the iron K{\ensuremath{\alpha}} line in NGC 7213}",
      journal = {\mnras},
         year = 2008,
        month = sep,
       volume = {389},
       number = {1},
        pages = {L52-L56},
          doi = {10.1111/j.1745-3933.2008.00521.x},
archivePrefix = {arXiv},
       eprint = {0806.3876},
 primaryClass = {astro-ph},
       adsurl = {https://ui.adsabs.harvard.edu/abs/2008MNRAS.389L..52B}
}

@ARTICLE{Boller96,
       author = {{Boller}, T. and {Brandt}, W.~N. and {Fink}, H.},
        title = "{Soft X-ray properties of narrow-line Seyfert 1 galaxies.}",
      journal = {\aap},
         year = 1996,
        month = jan,
       volume = {305},
        pages = {53},
          doi = {10.48550/arXiv.astro-ph/9504093},
archivePrefix = {arXiv},
       eprint = {astro-ph/9504093},
 primaryClass = {astro-ph},
       adsurl = {https://ui.adsabs.harvard.edu/abs/1996A&A...305...53B}
}

@ARTICLE{Boller07,
       author = {{Boller}, Th. and {Balestra}, I. and {Kollatschny}, W.},
        title = "{XMM-Newton observation of Mrk 110}",
      journal = {\aap},
         year = 2007,
        month = apr,
       volume = {465},
       number = {1},
        pages = {87-93},
          doi = {10.1051/0004-6361:20066343},
archivePrefix = {arXiv},
       eprint = {astro-ph/0612491},
 primaryClass = {astro-ph},
       adsurl = {https://ui.adsabs.harvard.edu/abs/2007A&A...465...87B}
}

@ARTICLE{Boroson92,
       author = {{Boroson}, Todd A. and {Green}, Richard F.},
        title = "{The Emission-Line Properties of Low-Redshift Quasi-stellar Objects}",
      journal = {\apjs},
         year = 1992,
        month = may,
       volume = {80},
        pages = {109},
          doi = {10.1086/191661},
       adsurl = {https://ui.adsabs.harvard.edu/abs/1992ApJS...80..109B}
}

@ARTICLE{Bower06,
       author = {{Bower}, R.~G. and {Benson}, A.~J. and {Malbon}, R. and {Helly}, J.~C. and {Frenk}, C.~S. and {Baugh}, C.~M. and {Cole}, S. and {Lacey}, C.~G.},
        title = "{Breaking the hierarchy of galaxy formation}",
      journal = {\mnras},
         year = 2006,
        month = aug,
       volume = {370},
       number = {2},
        pages = {645-655},
          doi = {10.1111/j.1365-2966.2006.10519.x},
archivePrefix = {arXiv},
       eprint = {astro-ph/0511338},
 primaryClass = {astro-ph},
       adsurl = {https://ui.adsabs.harvard.edu/abs/2006MNRAS.370..645B}
}

@ARTICLE{Brandt97,
       author = {{Brandt}, W.~N. and {Mathur}, S. and {Elvis}, M.},
        title = "{A comparison of the hard ASCA spectral slopes of broad- and narrow-line Seyfert 1 galaxies}",
      journal = {\mnras},
         year = 1997,
        month = mar,
       volume = {285},
       number = {3},
        pages = {L25-L30},
          doi = {10.1093/mnras/285.3.L25},
archivePrefix = {arXiv},
       eprint = {astro-ph/9703100},
 primaryClass = {astro-ph},
       adsurl = {https://ui.adsabs.harvard.edu/abs/1997MNRAS.285L..25B}
}

@ARTICLE{Cash79,
       author = {{Cash}, W.},
        title = "{Parameter estimation in astronomy through application of the likelihood ratio.}",
      journal = {\apj},
         year = 1979,
        month = mar,
       volume = {228},
        pages = {939-947},
          doi = {10.1086/156922},
       adsurl = {https://ui.adsabs.harvard.edu/abs/1979ApJ...228..939C}
}

@ARTICLE{Chartas09,
       author = {{Chartas}, G. and {Saez}, C. and {Brandt}, W.~N. and {Giustini}, M. and {Garmire}, G.~P.},
        title = "{Confirmation of and Variable Energy Injection by a Near-Relativistic Outflow in APM 08279+5255}",
      journal = {\apj},
         year = 2009,
        month = nov,
       volume = {706},
       number = {1},
        pages = {644-656},
          doi = {10.1088/0004-637X/706/1/64410.48550/arXiv.0910.0021},
archivePrefix = {arXiv},
       eprint = {0910.0021},
 primaryClass = {astro-ph.HE},
       adsurl = {https://ui.adsabs.harvard.edu/abs/2009ApJ...706..644C}
}

@ARTICLE{DiMatteo05,
       author = {{Di Matteo}, Tiziana and {Springel}, Volker and {Hernquist}, Lars},
        title = "{Energy input from quasars regulates the growth and activity of black holes and their host galaxies}",
      journal = {\nat},
         year = 2005,
        month = feb,
       volume = {433},
       number = {7026},
        pages = {604-607},
          doi = {10.1038/nature03335},
archivePrefix = {arXiv},
       eprint = {astro-ph/0502199},
 primaryClass = {astro-ph},
       adsurl = {https://ui.adsabs.harvard.edu/abs/2005Natur.433..604D}
}

@ARTICLE{Duras20,
       author = {{Duras}, F. and {Bongiorno}, A. and {Ricci}, F. and {Piconcelli}, E. and {Shankar}, F. and {Lusso}, E. and {Bianchi}, S. and {Fiore}, F. and {Maiolino}, R. and {Marconi}, A. and {Onori}, F. and {Sani}, E. and {Schneider}, R. and {Vignali}, C. and {La Franca}, F.},
        title = "{Universal bolometric corrections for active galactic nuclei over seven luminosity decades}",
      journal = {\aap},
         year = 2020,
        month = apr,
       volume = {636},
          eid = {A73},
        pages = {A73},
          doi = {10.1051/0004-6361/201936817},
archivePrefix = {arXiv},
       eprint = {2001.09984},
 primaryClass = {astro-ph.GA},
       adsurl = {https://ui.adsabs.harvard.edu/abs/2020A&A...636A..73D}
}

@ARTICLE{Feruglio15,
       author = {{Feruglio}, C. and {Fiore}, F. and {Carniani}, S. and {Piconcelli}, E. and {Zappacosta}, L. and {Bongiorno}, A. and {Cicone}, C. and {Maiolino}, R. and {Marconi}, A. and {Menci}, N. and {Puccetti}, S. and {Veilleux}, S.},
        title = "{The multi-phase winds of Markarian 231: from the hot, nuclear, ultra-fast wind to the galaxy-scale, molecular outflow}",
      journal = {\aap},
         year = 2015,
        month = nov,
       volume = {583},
          eid = {A99},
        pages = {A99},
          doi = {10.1051/0004-6361/201526020},
archivePrefix = {arXiv},
       eprint = {1503.01481},
 primaryClass = {astro-ph.GA},
       adsurl = {https://ui.adsabs.harvard.edu/abs/2015A&A...583A..99F}
}

@ARTICLE{Fields05,
       author = {{Fields}, Dale L. and {Mathur}, Smita and {Pogge}, Richard W. and {Nicastro}, Fabrizio and {Komossa}, Stefanie},
        title = "{Supersolar N/C in the Narrow-Line Seyfert 1 Galaxy Markarian 1044}",
      journal = {\apj},
         year = 2005,
        month = feb,
       volume = {620},
       number = {1},
        pages = {183-190},
          doi = {10.1086/426805},
archivePrefix = {arXiv},
       eprint = {astro-ph/0406042},
 primaryClass = {astro-ph},
       adsurl = {https://ui.adsabs.harvard.edu/abs/2005ApJ...620..183F}
}

@ARTICLE{Fields055,
       author = {{Fields}, Dale L. and {Mathur}, Smita and {Pogge}, Richard W. and {Nicastro}, Fabrizio and {Komossa}, Stefanie and {Krongold}, Yair},
        title = "{Supersolar Metallicity in the NLS1 Galaxy Markarian 1044}",
      journal = {\apj},
         year = 2005,
        month = dec,
       volume = {634},
       number = {2},
        pages = {928-938},
          doi = {10.1086/497121},
       adsurl = {https://ui.adsabs.harvard.edu/abs/2005ApJ...634..928F}
}

@ARTICLE{Gallo18,
       author = {{Gallo}, L.~C. and {Blue}, D.~M. and {Grupe}, D. and {Komossa}, S. and {Wilkins}, D.~R.},
        title = "{Eleven years of monitoring the Seyfert 1 Mrk 335 with Swift: Characterizing the X-ray and UV/optical variability}",
      journal = {\mnras},
         year = 2018,
        month = aug,
       volume = {478},
       number = {2},
        pages = {2557-2568},
          doi = {10.1093/mnras/sty1134},
archivePrefix = {arXiv},
       eprint = {1805.00300},
 primaryClass = {astro-ph.HE},
       adsurl = {https://ui.adsabs.harvard.edu/abs/2018MNRAS.478.2557G}
}

@ARTICLE{Gofford13,
       author = {{Gofford}, Jason and {Reeves}, James N. and {Tombesi}, Francesco and {Braito}, Valentina and {Turner}, T. Jane and {Miller}, Lance and {Cappi}, Massimo},
        title = "{The Suzaku view of highly ionized outflows in AGN - I. Statistical detection and global absorber properties}",
      journal = {\mnras},
         year = 2013,
        month = mar,
       volume = {430},
       number = {1},
        pages = {60-80},
          doi = {10.1093/mnras/sts48110.48550/arXiv.1211.5810},
archivePrefix = {arXiv},
       eprint = {1211.5810},
 primaryClass = {astro-ph.HE},
       adsurl = {https://ui.adsabs.harvard.edu/abs/2013MNRAS.430...60G}
}

@ARTICLE{Grupe06,
       author = {{Grupe}, D. and {Thomas}, H. -C. and {Beuermann}, K.},
        title = "{X-ray variability in a complete sample of Soft X-ray selected AGN}",
      journal = {\aap},
         year = 2001,
        month = feb,
       volume = {367},
        pages = {470-486},
          doi = {10.1051/0004-6361:20000429},
archivePrefix = {arXiv},
       eprint = {astro-ph/0012100},
 primaryClass = {astro-ph},
       adsurl = {https://ui.adsabs.harvard.edu/abs/2001A&A...367..470G}
}

@ARTICLE{Grupe04,
       author = {{Grupe}, Dirk and {Mathur}, Smita},
        title = "{M$_{BH}$-{\ensuremath{\sigma}} Relation for a Complete Sample of Soft X-Ray-selected Active Galactic Nuclei}",
      journal = {\apjl},
         year = 2004,
        month = may,
       volume = {606},
       number = {1},
        pages = {L41-L44},
          doi = {10.1086/420975},
archivePrefix = {arXiv},
       eprint = {astro-ph/0312390},
 primaryClass = {astro-ph},
       adsurl = {https://ui.adsabs.harvard.edu/abs/2004ApJ...606L..41G}
}

@ARTICLE{Gupta13,
       author = {{Gupta}, A. and {Mathur}, S. and {Krongold}, Y. and {Nicastro}, F.},
        title = "{Discovery of Relativistic Outflow in the Seyfert Galaxy Ark 564}",
      journal = {\apj},
         year = 2013,
        month = jul,
       volume = {772},
       number = {1},
          eid = {66},
        pages = {66},
          doi = {10.1088/0004-637X/772/1/66},
archivePrefix = {arXiv},
       eprint = {1301.6139},
 primaryClass = {astro-ph.HE},
       adsurl = {https://ui.adsabs.harvard.edu/abs/2013ApJ...772...66G}
}

@ARTICLE{Gupta15,
       author = {{Gupta}, A. and {Mathur}, S. and {Krongold}, Y.},
        title = "{Detection of High Velocity Outflows in the Seyfert 1 Galaxy Mrk 590}",
      journal = {\apj},
         year = 2015,
        month = jan,
       volume = {798},
       number = {1},
          eid = {4},
        pages = {4},
          doi = {10.1088/0004-637X/798/1/4},
archivePrefix = {arXiv},
       eprint = {1406.5968},
 primaryClass = {astro-ph.HE},
       adsurl = {https://ui.adsabs.harvard.edu/abs/2015ApJ...798....4G}
}

@ARTICLE{Harrison14,
       author = {{Harrison}, C.~M. and {Alexander}, D.~M. and {Mullaney}, J.~R. and {Swinbank}, A.~M.},
        title = "{Kiloparsec-scale outflows are prevalent among luminous AGN: outflows and feedback in the context of the overall AGN population}",
      journal = {\mnras},
         year = 2014,
        month = jul,
       volume = {441},
       number = {4},
        pages = {3306-3347},
          doi = {10.1093/mnras/stu515},
archivePrefix = {arXiv},
       eprint = {1403.3086},
 primaryClass = {astro-ph.GA},
       adsurl = {https://ui.adsabs.harvard.edu/abs/2014MNRAS.441.3306H}
}

@ARTICLE{Harrison18,
       author = {{Harrison}, C.~M. and {Costa}, T. and {Tadhunter}, C.~N. and {Fl{\"u}tsch}, A. and {Kakkad}, D. and {Perna}, M. and {Vietri}, G.},
        title = "{AGN outflows and feedback twenty years on}",
      journal = {Nature Astronomy},
         year = 2018,
        month = feb,
       volume = {2},
        pages = {198-205},
          doi = {10.1038/s41550-018-0403-6},
archivePrefix = {arXiv},
       eprint = {1802.10306},
 primaryClass = {astro-ph.GA},
       adsurl = {https://ui.adsabs.harvard.edu/abs/2018NatAs...2..198H}
}

@ARTICLE{Hebbar19,
       author = {{Hebbar}, Pavan R. and {Heinke}, Craig O. and {Sivakoff}, Gregory R. and {Shaw}, Aarran W.},
        title = "{X-ray spectroscopy of the candidate AGNs in Henize 2-10 and NGC 4178: likely supernova remnants}",
      journal = {\mnras},
         year = 2019,
        month = jun,
       volume = {485},
       number = {4},
        pages = {5604-5615},
          doi = {10.1093/mnras/stz553},
archivePrefix = {arXiv},
       eprint = {1902.08293},
 primaryClass = {astro-ph.HE},
       adsurl = {https://ui.adsabs.harvard.edu/abs/2019MNRAS.485.5604H}
}

@ARTICLE{Hopkins10,
       author = {{Hopkins}, Philip F. and {Elvis}, Martin},
        title = "{Quasar feedback: more bang for your buck}",
      journal = {\mnras},
         year = 2010,
        month = jan,
       volume = {401},
       number = {1},
        pages = {7-14},
          doi = {10.1111/j.1365-2966.2009.15643.x},
archivePrefix = {arXiv},
       eprint = {0904.0649},
 primaryClass = {astro-ph.CO},
       adsurl = {https://ui.adsabs.harvard.edu/abs/2010MNRAS.401....7H}
}

@ARTICLE{Juranova24,
       author = {{Jur{\'a}{\v{n}}ov{\'a}}, A. and {Costantini}, E. and {Di Gesu}, L. and {Ebrero}, J. and {Kaastra}, J. and {Korista}, K. and {Kriss}, G.~A. and {Mehdipour}, M. and {Piconcelli}, E. and {Rogantini}, D.},
        title = "{Study of the optical to X-ray broad emission lines of Mrk 110}",
      journal = {\aap},
         year = 2024,
        month = jan,
       volume = {681},
          eid = {A101},
        pages = {A101},
          doi = {10.1051/0004-6361/202348067},
archivePrefix = {arXiv},
       eprint = {2311.03459},
 primaryClass = {astro-ph.GA},
       adsurl = {https://ui.adsabs.harvard.edu/abs/2024A&A...681A.101J}
}

@ARTICLE{Karlberla05,
       author = {{Kalberla}, P.~M.~W. and {Burton}, W.~B. and {Hartmann}, Dap and {Arnal}, E.~M. and {Bajaja}, E. and {Morras}, R. and {P{\"o}ppel}, W.~G.~L.},
        title = "{The Leiden/Argentine/Bonn (LAB) Survey of Galactic HI. Final data release of the combined LDS and IAR surveys with improved stray-radiation corrections}",
      journal = {\aap},
         year = 2005,
        month = sep,
       volume = {440},
       number = {2},
        pages = {775-782},
          doi = {10.1051/0004-6361:20041864},
archivePrefix = {arXiv},
       eprint = {astro-ph/0504140},
 primaryClass = {astro-ph},
       adsurl = {https://ui.adsabs.harvard.edu/abs/2005A&A...440..775K}
}

@ARTICLE{King10,
       author = {{King}, A.~R.},
        title = "{Black hole outflows}",
      journal = {\mnras},
         year = 2010,
        month = mar,
       volume = {402},
       number = {3},
        pages = {1516-1522},
          doi = {10.1111/j.1365-2966.2009.16013.x10.48550/arXiv.0911.1639},
archivePrefix = {arXiv},
       eprint = {0911.1639},
 primaryClass = {astro-ph.CO},
       adsurl = {https://ui.adsabs.harvard.edu/abs/2010MNRAS.402.1516K}
}

@ARTICLE{Kollatschny03,
       author = {{Kollatschny}, W.},
        title = "{Accretion disk wind in the AGN broad-line region: Spectroscopically resolved line profile variations in Mrk 110}",
      journal = {\aap},
         year = 2003,
        month = aug,
       volume = {407},
        pages = {461-472},
          doi = {10.1051/0004-6361:20030928},
archivePrefix = {arXiv},
       eprint = {astro-ph/0306389},
 primaryClass = {astro-ph},
       adsurl = {https://ui.adsabs.harvard.edu/abs/2003A&A...407..461K}
}

@ARTICLE{Kollatschny06,
       author = {{Kollatschny}, W. and {Zetzl}, M. and {Dietrich}, M.},
        title = "{Spectral line variability amplitudes in active galactic nuclei}",
      journal = {\aap},
         year = 2006,
        month = aug,
       volume = {454},
       number = {2},
        pages = {459-472},
          doi = {10.1051/0004-6361:20054357},
       adsurl = {https://ui.adsabs.harvard.edu/abs/2006A&A...454..459K}
}

@ARTICLE{Kosec18,
       author = {{Kosec}, P. and {Buisson}, D.~J.~K. and {Parker}, M.~L. and {Pinto}, C. and {Fabian}, A.~C. and {Walton}, D.~J.},
        title = "{A stratified ultrafast outflow in 1H0707-495?}",
      journal = {\mnras},
         year = 2018,
        month = nov,
       volume = {481},
       number = {1},
        pages = {947-953},
          doi = {10.1093/mnras/sty2342},
archivePrefix = {arXiv},
       eprint = {1808.08204},
 primaryClass = {astro-ph.HE},
       adsurl = {https://ui.adsabs.harvard.edu/abs/2018MNRAS.481..947K}
}

@ARTICLE{Kormendy13,
       author = {{Kormendy}, John and {Ho}, Luis C.},
        title = "{Coevolution (Or Not) of Supermassive Black Holes and Host Galaxies}",
      journal = {\araa},
         year = 2013,
        month = aug,
       volume = {51},
       number = {1},
        pages = {511-653},
          doi = {10.1146/annurev-astro-082708-101811},
archivePrefix = {arXiv},
       eprint = {1304.7762},
 primaryClass = {astro-ph.CO},
       adsurl = {https://ui.adsabs.harvard.edu/abs/2013ARA&A..51..511K}
}

@ARTICLE{Krongold21,
       author = {{Krongold}, Y. and {Longinotti}, A.~L. and {Santos-Lle{\'o}}, M. and {Mathur}, S. and {Peterson}, B.~M. and {Nicastro}, F. and {Gupta}, A. and {Rodr{\'\i}guez-Pascual}, P. and {El{\'\i}as-Ch{\'a}vez}, M.},
        title = "{Detection of a Multiphase Ultrafast Wind in the Narrow-line Seyfert 1 Galaxy Mrk 1044}",
      journal = {\apj},
         year = 2021,
        month = aug,
       volume = {917},
       number = {1},
          eid = {39},
        pages = {39},
          doi = {10.3847/1538-4357/ac097710.48550/arXiv.2106.04725},
archivePrefix = {arXiv},
       eprint = {2106.04725},
 primaryClass = {astro-ph.GA},
       adsurl = {https://ui.adsabs.harvard.edu/abs/2021ApJ...917...39K}
}

@ARTICLE{Longinotti07,
       author = {{Longinotti}, A.~L. and {Bianchi}, S. and {Santos-Lleo}, M. and {Rodr{\'\i}guez-Pascual}, P. and {Guainazzi}, M. and {Cardaci}, M. and {Pollock}, A.~M.~T.},
        title = "{An X-ray look at the Seyfert 1 Galaxy Mrk 590: XMM-Newton and Chandra reveal complexity in circumnuclear gas}",
      journal = {\aap},
         year = 2007,
        month = jul,
       volume = {470},
       number = {1},
        pages = {73-81},
          doi = {10.1051/0004-6361:20066248},
archivePrefix = {arXiv},
       eprint = {astro-ph/0701180},
 primaryClass = {astro-ph},
       adsurl = {https://ui.adsabs.harvard.edu/abs/2007A&A...470...73L}
}

@ARTICLE{Longinotti15,
       author = {{Longinotti}, A.~L. and {Krongold}, Y. and {Guainazzi}, M. and {Giroletti}, M. and {Panessa}, F. and {Costantini}, E. and {Santos-Lleo}, M. and {Rodriguez-Pascual}, P.},
        title = "{X-Ray High-resolution Spectroscopy Reveals Feedback in a Seyfert Galaxy from an Ultra-fast Wind with Complex Ionization and Velocity Structure}",
      journal = {\apjl},
         year = 2015,
        month = nov,
       volume = {813},
       number = {2},
          eid = {L39},
        pages = {L39},
          doi = {10.1088/2041-8205/813/2/L3910.48550/arXiv.1511.01165},
archivePrefix = {arXiv},
       eprint = {1511.01165},
 primaryClass = {astro-ph.GA},
       adsurl = {https://ui.adsabs.harvard.edu/abs/2015ApJ...813L..39L}
}

@ARTICLE{Longinotti18,
       author = {{Longinotti}, A.~L. and {Vega}, O. and {Krongold}, Y. and {Aretxaga}, I. and {Yun}, M. and {Chavushyan}, V. and {Feruglio}, C. and {G{\'o}mez-Ruiz}, A. and {Monta{\~n}a}, A. and {Le{\'o}n-Tavares}, J. and {Olgu{\'\i}n-Iglesias}, A. and {Giroletti}, M. and {Guainazzi}, M. and {Kotilainen}, J. and {Panessa}, F. and {Zapata}, L.~A. and {Cruz-Gonzalez}, I. and {Pati{\~n}o-{\'A}lvarez}, V.~M. and {Rosa-Gonzalez}, D. and {Carrami{\~n}ana}, A. and {Carrasco}, L. and {Costantini}, E. and {Dultzin}, D. and {Guichard}, J. and {Puerari}, I. and {Santos-Lleo}, M.},
        title = "{Early Science with the Large Millimeter Telescope: An Energy-driven Wind Revealed by Massive Molecular and Fast X-Ray Outflows in the Seyfert Galaxy IRAS 17020+4544}",
      journal = {\apjl},
         year = 2018,
        month = nov,
       volume = {867},
       number = {1},
          eid = {L11},
        pages = {L11},
          doi = {10.3847/2041-8213/aae5fd},
archivePrefix = {arXiv},
       eprint = {1810.01941},
 primaryClass = {astro-ph.GA},
       adsurl = {https://ui.adsabs.harvard.edu/abs/2018ApJ...867L..11L}
}

@ARTICLE{Longinotti23,
       author = {{Longinotti}, Anna Lia and {Salom{\'e}}, Q. and {Feruglio}, C. and {Krongold}, Y. and {Garc{\'\i}a-Burillo}, S. and {Giroletti}, M. and {Panessa}, F. and {Stanghellini}, C. and {Vega}, O. and {Pati{\~n}o-{\'A}lvarez}, V.~M. and {Chavushyan}, V. and {El{\'\i}as-Chavez}, M. and {Robleto-Or{\'u}s}, A.},
        title = "{NOEMA spatially resolved view of the multiphase outflow in IRAS17020+4544: a shocked wind in action?}",
      journal = {\mnras},
         year = 2023,
        month = may,
       volume = {521},
       number = {2},
        pages = {2134-2148},
          doi = {10.1093/mnras/stad540},
archivePrefix = {arXiv},
       eprint = {2302.08165},
 primaryClass = {astro-ph.GA},
       adsurl = {https://ui.adsabs.harvard.edu/abs/2023MNRAS.521.2134L}
}

@ARTICLE{Luminari21,
       author = {{Luminari}, A. and {Nicastro}, F. and {Elvis}, M. and {Piconcelli}, E. and {Tombesi}, F. and {Zappacosta}, L. and {Fiore}, F.},
        title = "{Speed limits for radiation-driven SMBH winds}",
      journal = {\aap},
         year = 2021,
        month = feb,
       volume = {646},
          eid = {A111},
        pages = {A111},
          doi = {10.1051/0004-6361/202039396},
archivePrefix = {arXiv},
       eprint = {2012.07877},
 primaryClass = {astro-ph.HE},
       adsurl = {https://ui.adsabs.harvard.edu/abs/2021A&A...646A.111L}
}

@ARTICLE{Marasco20,
       author = {{Marasco}, A. and {Cresci}, G. and {Nardini}, E. and {Mannucci}, F. and {Marconi}, A. and {Tozzi}, P. and {Tozzi}, G. and {Amiri}, A. and {Venturi}, G. and {Piconcelli}, E. and {Lanzuisi}, G. and {Tombesi}, F. and {Mingozzi}, M. and {Perna}, M. and {Carniani}, S. and {Brusa}, M. and {di Serego Alighieri}, S.},
        title = "{Galaxy-scale ionised winds driven by ultra-fast outflows in two nearby quasars}",
      journal = {\aap},
         year = 2020,
        month = dec,
       volume = {644},
          eid = {A15},
        pages = {A15},
          doi = {10.1051/0004-6361/202038889},
archivePrefix = {arXiv},
       eprint = {2009.11294},
 primaryClass = {astro-ph.GA},
       adsurl = {https://ui.adsabs.harvard.edu/abs/2020A&A...644A..15M}
}

@ARTICLE{Mathur05,
       author = {{Mathur}, Smita and {Grupe}, Dirk},
        title = "{The Locus of Highly Accreting Active Galactic Nuclei on the M$_{BH}$-{\ensuremath{\sigma}} Plane: Selections, Limitations, and Implications}",
      journal = {\apj},
         year = 2005,
        month = nov,
       volume = {633},
       number = {2},
        pages = {688-692},
          doi = {10.1086/491613},
archivePrefix = {arXiv},
       eprint = {astro-ph/0507624},
 primaryClass = {astro-ph},
       adsurl = {https://ui.adsabs.harvard.edu/abs/2005ApJ...633..688M}
}

@ARTICLE{Mathur055,
       author = {{Mathur}, S. and {Grupe}, D.},
        title = "{Black hole growth by accretion}",
      journal = {\aap},
         year = 2005,
        month = mar,
       volume = {432},
       number = {2},
        pages = {463-466},
          doi = {10.1051/0004-6361:20041717},
archivePrefix = {arXiv},
       eprint = {astro-ph/0407512},
 primaryClass = {astro-ph},
       adsurl = {https://ui.adsabs.harvard.edu/abs/2005A&A...432..463M}
}

@ARTICLE{Mathur12,
       author = {{Mathur}, Smita and {Fields}, Dale and {Peterson}, Bradley M. and {Grupe}, Dirk},
        title = "{Supermassive Black Holes, Pseudobulges, and the Narrow-line Seyfert 1 Galaxies}",
      journal = {\apj},
         year = 2012,
        month = aug,
       volume = {754},
       number = {2},
          eid = {146},
        pages = {146},
          doi = {10.1088/0004-637X/754/2/146},
archivePrefix = {arXiv},
       eprint = {1102.0537},
 primaryClass = {astro-ph.CO},
       adsurl = {https://ui.adsabs.harvard.edu/abs/2012ApJ...754..146M}
}

@ARTICLE{Mathur00,
       author = {{Mathur}, Smita},
        title = "{Narrow-line Seyfert 1 galaxies and the evolution of galaxies and active galaxies}",
      journal = {\mnras},
         year = 2000,
        month = jun,
       volume = {314},
       number = {4},
        pages = {L17-L20},
          doi = {10.1046/j.1365-8711.2000.03530.x},
archivePrefix = {arXiv},
       eprint = {astro-ph/0003111},
 primaryClass = {astro-ph},
       adsurl = {https://ui.adsabs.harvard.edu/abs/2000MNRAS.314L..17M}
}

@ARTICLE{Matzeu17,
       author = {{Matzeu}, G.~A. and {Reeves}, J.~N. and {Braito}, V. and {Nardini}, E. and {McLaughlin}, D.~E. and {Lobban}, A.~P. and {Tombesi}, F. and {Costa}, M.~T.},
        title = "{Evidence for a radiatively driven disc-wind in PDS 456?}",
      journal = {\mnras},
         year = 2017,
        month = nov,
       volume = {472},
       number = {1},
        pages = {L15-L19},
          doi = {10.1093/mnrasl/slx129},
archivePrefix = {arXiv},
       eprint = {1708.03546},
 primaryClass = {astro-ph.HE},
       adsurl = {https://ui.adsabs.harvard.edu/abs/2017MNRAS.472L..15M}
}

@ARTICLE{Matzeu23,
       author = {{Matzeu}, G.~A. and {Brusa}, M. and {Lanzuisi}, G. and {Dadina}, M. and {Bianchi}, S. and {Kriss}, G. and {Mehdipour}, M. and {Nardini}, E. and {Chartas}, G. and {Middei}, R. and {Piconcelli}, E. and {Gianolli}, V. and {Comastri}, A. and {Longinotti}, A.~L. and {Krongold}, Y. and {Ricci}, F. and {Petrucci}, P.~O. and {Tombesi}, F. and {Luminari}, A. and {Zappacosta}, L. and {Miniutti}, G. and {Gaspari}, M. and {Behar}, E. and {Bischetti}, M. and {Mathur}, S. and {Perna}, M. and {Giustini}, M. and {Grandi}, P. and {Torresi}, E. and {Vignali}, C. and {Bruni}, G. and {Cappi}, M. and {Costantini}, E. and {Cresci}, G. and {De Marco}, B. and {De Rosa}, A. and {Gilli}, R. and {Guainazzi}, M. and {Kaastra}, J. and {Kraemer}, S. and {La Franca}, F. and {Marconi}, A. and {Panessa}, F. and {Ponti}, G. and {Proga}, D. and {Ursini}, F. and {Baldini}, P. and {Fiore}, F. and {King}, A.~R. and {Maiolino}, R. and {Matt}, G. and {Merloni}, A.},
        title = "{Supermassive Black Hole Winds in X-rays: SUBWAYS. I. Ultra-fast outflows in quasars beyond the local Universe}",
      journal = {\aap},
         year = 2023,
        month = feb,
       volume = {670},
          eid = {A182},
        pages = {A182},
          doi = {10.1051/0004-6361/202245036},
archivePrefix = {arXiv},
       eprint = {2212.02960},
 primaryClass = {astro-ph.HE},
       adsurl = {https://ui.adsabs.harvard.edu/abs/2023A&A...670A.182M}
}

@ARTICLE{Mizumoto26,
       author = {{Mizumoto}, Misaki and {Reeves}, James N. and {Braito}, Valentina and {Behar}, Ehud and {Done}, Chris and {Hagino}, Kouichi and {Kraemer}, Steven B. and {Matzeu}, Gabriele A. and {Noda}, Hirofumi and {Nomura}, Mariko and {Ogawa}, Shoji and {Ohsuga}, Ken and {Tanimoto}, Atsushi and {Turner}, Tracey J. and {Ueda}, Yoshihiro and {Yamada}, Satoshi and {Ganguly}, Sreeparna and {Somenzi}, Paolo},
        title = "{Resolving the Multiple Component Outflows in PG 1211+143. I. The Fe─K Absorption Structure and UFO Forest}",
      journal = {\apj},
         year = 2026,
        month = feb,
       volume = {997},
       number = {2},
          eid = {219},
        pages = {219},
          doi = {10.3847/1538-4357/ae2853},
archivePrefix = {arXiv},
       eprint = {2512.03533},
 primaryClass = {astro-ph.HE},
       adsurl = {https://ui.adsabs.harvard.edu/abs/2026ApJ...997..219M}
}

@ARTICLE{Osterbrock85,
       author = {{Osterbrock}, D.~E. and {Pogge}, R.~W.},
        title = "{The spectra of narrow-line Seyfert 1 galaxies.}",
      journal = {\apj},
         year = 1985,
        month = oct,
       volume = {297},
        pages = {166-176},
          doi = {10.1086/163513},
       adsurl = {https://ui.adsabs.harvard.edu/abs/1985ApJ...297..166O}
}

@ARTICLE{Panessa22,
       author = {{Panessa}, Francesca and {P{\'e}rez-Torres}, Miguel and {Hern{\'a}ndez-Garc{\'\i}a}, Lorena and {Casella}, Piergiorgio and {Giroletti}, Marcello and {Orienti}, Monica and {Baldi}, Ranieri D. and {Bassani}, Loredana and {Fiocchi}, Maria Teresa and {La Franca}, Fabio and {Malizia}, Angela and {McHardy}, Ian and {Nicastro}, Fabrizio and {Piro}, Luigi and {Vincentelli}, Federico and {Williams}, David R.~A. and {Ubertini}, Pietro},
        title = "{Daily variability at milli-arcsecond scales in the radio-quiet NLSy1 Mrk 110}",
      journal = {\mnras},
         year = 2022,
        month = feb,
       volume = {510},
       number = {1},
        pages = {718-724},
          doi = {10.1093/mnras/stab3426},
archivePrefix = {arXiv},
       eprint = {2111.12595},
 primaryClass = {astro-ph.GA},
       adsurl = {https://ui.adsabs.harvard.edu/abs/2022MNRAS.510..718P}
}

@ARTICLE{Parker17,
       author = {{Parker}, Michael L. and {Pinto}, Ciro and {Fabian}, Andrew C. and {Lohfink}, Anne and {Buisson}, Douglas J.~K. and {Alston}, William N. and {Kara}, Erin and {Cackett}, Edward M. and {Chiang}, Chia-Ying and {Dauser}, Thomas and {De Marco}, Barbara and {Gallo}, Luigi C. and {Garcia}, Javier and {Harrison}, Fiona A. and {King}, Ashley L. and {Middleton}, Matthew J. and {Miller}, Jon M. and {Miniutti}, Giovanni and {Reynolds}, Christopher S. and {Uttley}, Phil and {Vasudevan}, Ranjan and {Walton}, Dominic J. and {Wilkins}, Daniel R. and {Zoghbi}, Abderahmen},
        title = "{The response of relativistic outflowing gas to the inner accretion disk of a black hole}",
      journal = {\nat},
         year = 2017,
        month = mar,
       volume = {543},
       number = {7643},
        pages = {83-86},
          doi = {10.1038/nature21385},
archivePrefix = {arXiv},
       eprint = {1703.00071},
 primaryClass = {astro-ph.HE},
       adsurl = {https://ui.adsabs.harvard.edu/abs/2017Natur.543...83P}
}

@ARTICLE{Pinto17,
       author = {{Pinto}, C. and {Alston}, W. and {Soria}, R. and {Middleton}, M.~J. and {Walton}, D.~J. and {Sutton}, A.~D. and {Fabian}, A.~C. and {Earnshaw}, H. and {Urquhart}, R. and {Kara}, E. and {Roberts}, T.~P.},
        title = "{From ultraluminous X-ray sources to ultraluminous supersoft sources: NGC 55 ULX, the missing link}",
      journal = {\mnras},
         year = 2017,
        month = jul,
       volume = {468},
       number = {3},
        pages = {2865-2883},
          doi = {10.1093/mnras/stx641},
archivePrefix = {arXiv},
       eprint = {1612.05569},
 primaryClass = {astro-ph.HE},
       adsurl = {https://ui.adsabs.harvard.edu/abs/2017MNRAS.468.2865P}
}

@ARTICLE{Pinto18,
       author = {{Pinto}, C. and {Alston}, W. and {Parker}, M.~L. and {Fabian}, A.~C. and {Gallo}, L.~C. and {Buisson}, D.~J.~K. and {Walton}, D.~J. and {Kara}, E. and {Jiang}, J. and {Lohfink}, A. and {Reynolds}, C.~S.},
        title = "{Ultrafast outflows disappear in high-radiation fields}",
      journal = {\mnras},
         year = 2018,
        month = may,
       volume = {476},
       number = {1},
        pages = {1021-1035},
          doi = {10.1093/mnras/sty231},
archivePrefix = {arXiv},
       eprint = {1708.09422},
 primaryClass = {astro-ph.HE},
       adsurl = {https://ui.adsabs.harvard.edu/abs/2018MNRAS.476.1021P}
}

@ARTICLE{Porquet21,
       author = {{Porquet}, D. and {Reeves}, J.~N. and {Grosso}, N. and {Braito}, V. and {Lobban}, A.},
        title = "{The first simultaneous X-ray broadband view of Mrk 110 with XMM-Newton and NuSTAR}",
      journal = {\aap},
         year = 2021,
        month = oct,
       volume = {654},
          eid = {A89},
        pages = {A89},
          doi = {10.1051/0004-6361/202141577},
archivePrefix = {arXiv},
       eprint = {2107.06733},
 primaryClass = {astro-ph.HE},
       adsurl = {https://ui.adsabs.harvard.edu/abs/2021A&A...654A..89P}
}

@ARTICLE{Pounds11,
       author = {{Pounds}, K.~A. and {Vaughan}, S.},
        title = "{An extended XMM-Newton observation of the Seyfert galaxy NGC 4051 - I. Evidence for a shocked outflow}",
      journal = {\mnras},
         year = 2011,
        month = may,
       volume = {413},
       number = {2},
        pages = {1251-1263},
          doi = {10.1111/j.1365-2966.2011.18211.x},
archivePrefix = {arXiv},
       eprint = {1012.0998},
 primaryClass = {astro-ph.CO},
       adsurl = {https://ui.adsabs.harvard.edu/abs/2011MNRAS.413.1251P}
}

@ARTICLE{Pounds14,
       author = {{Pounds}, K.~A.},
        title = "{The soft X-ray spectrum of the luminous narrow line Seyfert galaxy PG 1211+143 - evidence for a second high-velocity outflow component}",
      journal = {\mnras},
         year = 2014,
        month = feb,
       volume = {437},
       number = {4},
        pages = {3221-3227},
          doi = {10.1093/mnras/stt211310.48550/arXiv.1310.2836},
archivePrefix = {arXiv},
       eprint = {1310.2836},
 primaryClass = {astro-ph.HE},
       adsurl = {https://ui.adsabs.harvard.edu/abs/2014MNRAS.437.3221P}
}

@ARTICLE{Rees84,
       author = {{Rees}, Martin J.},
        title = "{Black Hole Models for Active Galactic Nuclei}",
      journal = {\araa},
         year = 1984,
        month = jan,
       volume = {22},
        pages = {471-506},
          doi = {10.1146/annurev.aa.22.090184.002351},
       adsurl = {https://ui.adsabs.harvard.edu/abs/1984ARA&A..22..471R}
}

@ARTICLE{Reeves19,
       author = {{Reeves}, J.~N. and {Braito}, V.},
        title = "{A Momentum-conserving Accretion Disk Wind in the Narrow-line Seyfert 1 I Zwicky 1}",
      journal = {\apj},
         year = 2019,
        month = oct,
       volume = {884},
       number = {1},
          eid = {80},
        pages = {80},
          doi = {10.3847/1538-4357/ab41f9},
archivePrefix = {arXiv},
       eprint = {1909.05039},
 primaryClass = {astro-ph.GA},
       adsurl = {https://ui.adsabs.harvard.edu/abs/2019ApJ...884...80R}
}

@ARTICLE{Reeves21,
       author = {{Reeves}, J.~N. and {Porquet}, D. and {Braito}, V. and {Grosso}, N. and {Lobban}, A.},
        title = "{Variable oxygen emission from the accretion disk of Mrk 110}",
      journal = {\aap},
         year = 2021,
        month = may,
       volume = {649},
          eid = {L3},
        pages = {L3},
          doi = {10.1051/0004-6361/202140953},
archivePrefix = {arXiv},
       eprint = {2104.14584},
 primaryClass = {astro-ph.HE},
       adsurl = {https://ui.adsabs.harvard.edu/abs/2021A&A...649L...3R}
}

@ARTICLE{Reeves18,
       author = {{Reeves}, J.~N. and {Braito}, V. and {Nardini}, E. and {Hamann}, F. and {Chartas}, G. and {Lobban}, A.~P. and {O'Brien}, P.~T. and {Turner}, T.~J.},
        title = "{Resolving the X-Ray Obscuration in a Low-flux Observation of the Quasar PDS 456}",
      journal = {\apj},
         year = 2018,
        month = nov,
       volume = {867},
       number = {1},
          eid = {38},
        pages = {38},
          doi = {10.3847/1538-4357/aae30c},
archivePrefix = {arXiv},
       eprint = {1809.07164},
 primaryClass = {astro-ph.HE},
       adsurl = {https://ui.adsabs.harvard.edu/abs/2018ApJ...867...38R}
}

@ARTICLE{Sanfrutos18,
       author = {{Sanfrutos}, Mario and {Longinotti}, Anna Lia and {Krongold}, Yair and {Guainazzi}, Matteo and {Panessa}, Francesca},
        title = "{The Evolution of the Warm Absorber Reveals a Shocked Outflow in the Narrow Line Seyfert 1 Galaxy IRAS 17020+4544}",
      journal = {\apj},
         year = 2018,
        month = dec,
       volume = {868},
       number = {2},
          eid = {111},
        pages = {111},
          doi = {10.3847/1538-4357/aae923},
archivePrefix = {arXiv},
       eprint = {1810.07714},
 primaryClass = {astro-ph.GA},
       adsurl = {https://ui.adsabs.harvard.edu/abs/2018ApJ...868..111S}
}

@ARTICLE{Tombesi10,
       author = {{Tombesi}, F. and {Sambruna}, R.~M. and {Reeves}, J.~N. and {Braito}, V. and {Ballo}, L. and {Gofford}, J. and {Cappi}, M. and {Mushotzky}, R.~F.},
        title = "{Discovery of Ultra-fast Outflows in a Sample of Broad-line Radio Galaxies Observed with Suzaku}",
      journal = {\apj},
         year = 2010,
        month = aug,
       volume = {719},
       number = {1},
        pages = {700-715},
          doi = {10.1088/0004-637X/719/1/70010.48550/arXiv.1006.3536},
archivePrefix = {arXiv},
       eprint = {1006.3536},
 primaryClass = {astro-ph.HE},
       adsurl = {https://ui.adsabs.harvard.edu/abs/2010ApJ...719..700T}
}

@ARTICLE{Tombesi15,
       author = {{Tombesi}, F. and {Mel{\'e}ndez}, M. and {Veilleux}, S. and {Reeves}, J.~N. and {Gonz{\'a}lez-Alfonso}, E. and {Reynolds}, C.~S.},
        title = "{Wind from the black-hole accretion disk driving a molecular outflow in an active galaxy}",
      journal = {\nat},
         year = 2015,
        month = mar,
       volume = {519},
       number = {7544},
        pages = {436-438},
          doi = {10.1038/nature14261},
archivePrefix = {arXiv},
       eprint = {1501.07664},
 primaryClass = {astro-ph.HE},
       adsurl = {https://ui.adsabs.harvard.edu/abs/2015Natur.519..436T}
}

@ARTICLE{Wilms00,
       author = {{Wilms}, J. and {Allen}, A. and {McCray}, R.},
        title = "{On the Absorption of X-Rays in the Interstellar Medium}",
      journal = {\apj},
         year = 2000,
        month = oct,
       volume = {542},
       number = {2},
        pages = {914-924},
          doi = {10.1086/317016},
archivePrefix = {arXiv},
       eprint = {astro-ph/0008425},
 primaryClass = {astro-ph},
       adsurl = {https://ui.adsabs.harvard.edu/abs/2000ApJ...542..914W}
}

@ARTICLE{pico04,
   author = {{Piconcelli}, E. and {Jimenez-Bail{\' o}n}, E. and {Guainazzi}, M. and
	{Schartel}, N. and {Rodr{\'{\i}}guez-Pascual}, P.~M. and {Santos-Lle{\' o}}, M.
	},
    title = "{Evidence for a multizone warm absorber in the XMM-Newton spectrum of Markarian 304}",
  journal = {\mnras},
     year = 2004,
    month = jun,
   volume = 351,
    pages = {161-168},
   adsurl = {http://esoads.eso.org/cgi-bin/nph-bib_query?bibcode=2004MNRAS.351..161P&amp;db_key=AST}
}

@ARTICLE{nustar,
   author = {{Harrison}, F.~A. and {Craig}, W.~W. and {Christensen}, F.~E. and 
	{Hailey}, C.~J. and {Zhang}, W.~W. and {Boggs}, S.~E. and {Stern}, D. and 
	{Cook}, W.~R. and {Forster}, K. and {Giommi}, P. and {Grefenstette}, B.~W. and 
	{Kim}, Y. and {Kitaguchi}, T. and {Koglin}, J.~E. and {Madsen}, K.~K. and 
	{Mao}, P.~H. and {Miyasaka}, H. and {Mori}, K. and {Perri}, M. and 
	{Pivovaroff}, M.~J. and {Puccetti}, S. and {Rana}, V.~R. and 
	{Westergaard}, N.~J. and {Willis}, J. and {Zoglauer}, A. and 
	{An}, H. and {Bachetti}, M. and {Barri{\`e}re}, N.~M. and {Bellm}, E.~C. and 
	{Bhalerao}, V. and {Brejnholt}, N.~F. and {Fuerst}, F. and {Liebe}, C.~C. and 
	{Markwardt}, C.~B. and {Nynka}, M. and {Vogel}, J.~K. and {Walton}, D.~J. and 
	{Wik}, D.~R. and {Alexander}, D.~M. and {Cominsky}, L.~R. and 
	{Hornschemeier}, A.~E. and {Hornstrup}, A. and {Kaspi}, V.~M. and 
	{Madejski}, G.~M. and {Matt}, G. and {Molendi}, S. and {Smith}, D.~M. and 
	{Tomsick}, J.~A. and {Ajello}, M. and {Ballantyne}, D.~R. and 
	{Balokovi{\'c}}, M. and {Barret}, D. and {Bauer}, F.~E. and 
	{Blandford}, R.~D. and {Niel Brandt}, W. and {Brenneman}, L.~W. and 
	{Chiang}, J. and {Chakrabarty}, D. and {Chenevez}, J. and {Comastri}, A. and 
	{Dufour}, F. and {Elvis}, M. and {Fabian}, A.~C. and {Farrah}, D. and 
	{Fryer}, C.~L. and {Gotthelf}, E.~V. and {Grindlay}, J.~E. and 
	{Helfand}, D.~J. and {Krivonos}, R. and {Meier}, D.~L. and {Miller}, J.~M. and 
	{Natalucci}, L. and {Ogle}, P. and {Ofek}, E.~O. and {Ptak}, A. and 
	{Reynolds}, S.~P. and {Rigby}, J.~R. and {Tagliaferri}, G. and 
	{Thorsett}, S.~E. and {Treister}, E. and {Urry}, C.~M.},
    title = "{The Nuclear Spectroscopic Telescope Array (NuSTAR) High-energy X-Ray Mission}",
  journal = {\apj},
archivePrefix = "arXiv",
   eprint = {1301.7307},
 primaryClass = "astro-ph.IM",
     year = 2013,
    month = jun,
   volume = 770,
      eid = {103},
    pages = {103},
      doi = {10.1088/0004-637X/770/2/103},
   adsurl = {http://adsabs.harvard.edu/abs/2013ApJ...770..103H}
}

@ARTICLE{Watson07,
       author = {{Watson}, Linda C. and {Mathur}, Smita and {Grupe}, Dirk},
        title = "{Revisiting the Black Hole Masses of Soft X-Ray-Selected Active Galactic Nuclei}",
      journal = {\aj},
         year = 2007,
        month = jun,
       volume = {133},
       number = {6},
        pages = {2435-2441},
          doi = {10.1086/515439},
archivePrefix = {arXiv},
       eprint = {astro-ph/0611819},
 primaryClass = {astro-ph},
       adsurl = {https://ui.adsabs.harvard.edu/abs/2007AJ....133.2435W}
}

@ARTICLE{Xu24,
       author = {{Xu}, Yerong and {Pinto}, Ciro and {Rogantini}, Daniele and {Barret}, Didier and {Bianchi}, Stefano and {Guainazzi}, Matteo and {Ebrero}, Jacobo and {Alston}, William and {Kara}, Erin and {Cusumano}, Giancarlo},
        title = "{A systematic study of the ultra-fast outflow responses to luminosity variations in active galactic nuclei}",
      journal = {arXiv e-prints},
         year = 2024,
        month = may,
          eid = {arXiv:2405.07494},
        pages = {arXiv:2405.07494},
          doi = {10.48550/arXiv.2405.07494},
archivePrefix = {arXiv},
       eprint = {2405.07494},
 primaryClass = {astro-ph.HE},
       adsurl = {https://ui.adsabs.harvard.edu/abs/2024arXiv240507494X}
}

@ARTICLE{Xiang25,
       author = {{Xiang}, Xin and {Miller}, Jon M. and {Behar}, Ehud and {Boissay-Malaquin}, Rozenn and {Brenneman}, Laura and {Buhariwalla}, Margaret and {Byun}, Doyee and {Done}, Chris and {Gallo}, Luigi and {Gerolymatou}, Dimitra and {Hagen}, Scott and {Kaastra}, Jelle and {Paltani}, Stephane and {Porter}, Frederick S. and {Mushotzky}, Richard and {Noda}, Hirofumi and {Mehdipour}, Missagh and {Minezaki}, Takeo and {Tashiro}, Makoto and {Zoghbi}, Abderahmen},
        title = "{XRISM Spectroscopy of Accretion-driven Wind Feedback in NGC 4151}",
      journal = {\apjl},
         year = 2025,
        month = aug,
       volume = {988},
       number = {2},
          eid = {L54},
        pages = {L54},
          doi = {10.3847/2041-8213/adee9b},
archivePrefix = {arXiv},
       eprint = {2507.09210},
 primaryClass = {astro-ph.HE},
       adsurl = {https://ui.adsabs.harvard.edu/abs/2025ApJ...988L..54X}
}

@ARTICLE{XRI1365,
       author = {{Zaidouni}, Fatima and {Kara}, Erin and {Kosec}, Peter and {Behar}, Ehud and {Mushotzky}, Richard and {Koss}, Michael and {Jur{\'a}{\v{n}}ov{\'a}}, A. and {Kammoun}, Elias and {Brenneman}, Laura W. and {Chakraborty}, Joheen and {Ebisawa}, Ken and {Eckart}, Megan E. and {Fabian}, Andrew C. and {Fukazawa}, Yasushi and {Garc{\'\i}a}, Javier A. and {Gu}, Liyi and {Masterson}, Megan and {Ogawa}, Shoji and {Okajima}, Takashi and {Paltani}, St{\'e}phane and {Rogantini}, Daniele and {Terashima}, Yuichi and {Williams}, Brian J. and {Yamada}, Satoshi},
        title = "{XRISM Finds the Changing-look Active Galactic Nucleus NGC 1365 in an Extended Low State: A Dense, Highly Ionized Outflow Obscures the Central Source}",
      journal = {\apjl},
         year = 2026,
        month = jan,
       volume = {997},
       number = {1},
          eid = {L26},
        pages = {L26},
          doi = {10.3847/2041-8213/ae307d},
archivePrefix = {arXiv},
       eprint = {2601.00795},
 primaryClass = {astro-ph.HE},
       adsurl = {https://ui.adsabs.harvard.edu/abs/2026ApJ...997L..26Z}
}


\end{document}